\documentclass[12pt,manuscript]{aastex631}

\usepackage{graphics,graphicx}
\graphicspath{./}
\usepackage{natbib}

\newcommand{\src}{G1.9+0.3}

\catcode`\@=11
\newcommand{\gapprox}{\mathrel{\mathpalette\@versim>}}
\newcommand{\lapprox}{\mathrel{\mathpalette\@versim<}}
\newcommand{\propapprox}{\mathrel{\mathpalette\@versim\propto}}
\newcommand{\@versim}[2]
  {\lower3.1truept\vbox{\baselineskip0pt\lineskip0.5truept
  \ialign{$\m@th#1\hfil##\hfil$\crcr#2\crcr\sim\crcr}}}
\catcode`\@=12

\shorttitle{TYPE IA SUPERNOVA MODELS AND SNR G1.9+0.3}

\begin{document}

\title{Type Ia Supernova Models:  Asymmetric Remnants and Supernova Remnant G1.9+0.3}

\author{Alice Griffeth Stone}
\affiliation{Department of Physics, North Carolina State University, 
Raleigh, NC 27695-8202}

\author{Heather T.~Johnson}
\affiliation{Department of Physics, North Carolina State University, 
Raleigh, NC 27695-8202} 
\affiliation{Department of Physics, University of Texas,
2515 Speedway, C1600, Austin, TX 78712-1192}

\author{John M.~Blondin}
\affiliation{Department of Physics, North Carolina State University, 
Raleigh, NC 27695-8202} 

\author{Richard A.~Watson}
\affiliation{Department of Physics, North Carolina State University, 
Raleigh, NC 27695-8202} 

\author[0000-0002-2614-1106]{Kazimierz J.~Borkowski}
\affiliation{Department of Physics, North Carolina State University, 
Raleigh, NC 27695-8202} 

\author[0000-0003-0191-2477]{Carla Fr\"ohlich}
\affiliation{Department of Physics, North Carolina State University, 
Raleigh, NC 27695-8202} 

\author{Ivo R.~Seitenzahl}
\affiliation{School of Science, The University of New South Wales, Australian Defence Force Academy, Northcott Drive, Canberra, ACT 2600, Australia}

\author[0000-0002-5365-5444]{Stephen P. Reynolds}
\affiliation{Department of Physics, North Carolina State University, 
Raleigh, NC 27695-8202}

\begin{abstract}

The youngest Galactic supernova remnant \src, probably the result of a Type Ia supernova, shows surprising anomalies in the distribution of its ejecta in space and velocity.  In particular, high-velocity shocked iron is seen in several locations far from the remnant center, in some cases beyond prominent silicon and sulfur emission.  These asymmetries strongly suggest a highly asymmetric explosion.  We present high-resolution hydrodynamic simulations in two and three dimensions of the evolution from ages of 100 seconds to hundreds of years of two asymmetric Type Ia models, expanding into a uniform medium.  At the age of \src\ (about 100 years), our 2D model shows almost no iron shocked to become visible in X-rays.  Only in a much higher-density environment could significant iron be shocked, at which time the model's expansion speed is completely inconsistent with the observations of \src.  Our 3D model, evolving the most asymmetric of a suite of Type Ia SN models from Seitenzahl et al.~(2013), shows some features resembling \src.  We characterize its evolution with images of composition in three classes: C and O, intermediate-mass elements (IMEs), and iron-group elements (IGEs).  From ages of 13 to 1800 years, we follow the evolution of the highly asymmetric initial remnant as the explosion asymmetries decrease in relative strength to be replaced by asymmetries due to evolutionary hydrodynamic instabilities.  At an age of about 100 years, our 3D model has comparable shocked masses of C+O, IMEs, and IGEs, with about 0.03 $M_\odot$ each.  Evolutionary changes appear to be rapid enough that continued monitoring with the {\sl Chandra} X-ray Observatory may show significant variations.

\end{abstract}

\section{Introduction}
\label{intro}

Type Ia supernovae perform vital functions in the Universe: providing
the majority of iron, as well as significant quantities of other
high-$Z$ elements; serving as standardizable candles for cosmography;
accelerating cosmic rays with very fast shocks in low-density media.
Given that importance, our poor understanding of the progenitor
systems, the immediate surroundings, and the fundamental mechanisms of
the explosion is a serious embarrassment.  The general agreement that
SNe Ia result from exploding white dwarfs is now over 30 years old.
However, the decades of work since then have failed to resolve the
question of the ignition mechanism or the nature of the nuclear
burning front that disrupts the star.  Detonation, deflagration,
delayed detonation, pulsed delayed detonation, and gravitationally
confined detonation models are still discussed and considered as viable models.
The ignition of either a subsonic (deflagration) or a supersonic (detonation)
thermonuclear burning front is a very complex astrophysical problem, and
consequently initial conditions for both are often implemented in 
astrophysical simulations in an ad-hoc manner.
Even putting aside the fraught question of the nature of the
progenitor systems (degenerate or non-degenerate companion, that is,
single-degenerate [SD] or double-degenerate [DD] systems), tremendous
uncertainty surrounds the basic nature of the explosion.  See
\cite{hillebrandt00}, \cite{hillebrandt13}, and \cite{ruiter20} 
for recent reviews.

While the basic idea of a Chandrasekhar-mass white dwarf exploding
(somehow) attractively explains the general homogeneity of SN Ia light
curves and spectra, at least compared to core-collapse events,
evidence for SN Ia diversity continues to increase \citep{benetti05,taubenberger17}
from the seminal discovery of the luminosity-width relation
\citep{phillips93} to the range of subclasses proposed in the last few
years, including SN 2002cx-like, or SNe Iax \citep{jha06, foley13},
underluminous or SN1991bg-like \citep{li01}, and SN 2002ic-like or SN
Ia-CSM \citep{silverman13}, among others.  One promising avenue to
explain this diversity, especially of spectral properties near maximum
light, is viewing-angle dependence of the appearance of asymmetric
explosions \citep{maeda10b}.  Asymmetry in explosions can result from
off-center ignition \citep{kasen09,maeda10b} or
 for instance by gravitationally confined detonation
\citep[GCD;][]{plewa04}.

Asymmetries in the supernova events themselves produce various
signatures, such as expansion velocity gradients \citep{maeda10b} or
nonzero polarization \citep[e.g.,][]{wang07}.  These have been used to
constrain explosion models, though it is often the case that multiple
models can reproduce the same observations \citep{hillebrandt13}.
However, one avenue for investigation of the nature of Type Ia
supernovae has been historically underutilized: the detailed
examination of Type Ia supernova remnants (SNRs).  The historical Type
Ia SNRs Tycho, Kepler, and SN 1006 have all been observed with long
exposures by all functioning X-ray satellites, and important clues to
the Type Ia phenomenon have been obtained.  Spatially resolved
spectroscopy provides direct information on anisotropies obtained only
with difficulty, if at all, from observations of the supernovae
themselves.  Tycho has a light-echo spectrum indicating a completely
normal SN Ia \citep{krause08}, and its X-ray spectrum shows
substantial radial and azimuthal asymmetries that can be traced to the
initial SN ejecta distribution \citep{miceli15}.  SN 1006, far above
the Galactic plane, is expanding into quite low-density material and
is consequently less evolved than Tycho or Kepler.  One manifestation
of this is that very little shocked iron is observed in X-rays
\citep{yamaguchi08}, indicating that large-scale overturn has not
occurred to move iron to large radii.  However, even when unshocked
iron seen in absorption in the UV \citep{winkler05} is included, the
total iron mass seems anomalously low.  Kepler's SNR is sufficiently
peculiar that its classification was still in doubt until quite
recently, although evidence has mounted for years pointing to its
thermonuclear nature (see \cite{reynolds07} for a summary of the
arguments).  But Kepler also shows unmistakable signs of interaction
with a circumstellar medium (CSM) modified by the progenitor system
\citep{burkey13,chiotellis13,kasuga21}, indicating a single-degenerate origin.
One additional remnant of probable Type Ia origin is RCW 86
\citep{williams11}, which shows evidence that the explosion took place
off-center in a wind-blown cavity.  More recently, \cite{seitenzahl19} 
detected coronal emission lines from iron in the young Type Ia SNRs 0509$-$67.5
and 0519$-$69.0 in the LMC, originating from the nonradiative reverse shock.
These observations allow discrimination among models of differing progenitor mass.

Multidimensional models of SNe Ia are now within reach of modern
computational resources \citep{hillebrandt13}, and modelers have
produced families of such models under various assumptions about
number and location of ignition points and other variables
\citep{kasen09,seitenzahl13}.  These explosion models are sometimes
used to generate predicted light curves and spectra 
\citep[e.g.,][]{sim13, fink14, noebauer17}. Since SN Ia
progenitors have such low masses, the explosions become ballistic
(that is, unaffected by pressure forces) in times of order 10 -- 100
s, and typical explosion models easily reach this stage.  This is
important, because subsequent evolution can then be described with
these models as initial conditions for pure hydrodynamic codes,
following the evolution for hundreds or thousands of years.  Such
calculations then open a whole new area of potential confrontation
between models and observations: the detailed comparison of
supernova-remnant properties with theoretical predictions.  This area
holds great promise for advancing our understanding of the Type Ia
phenomenon.

In recent years, several groups have considered the supernova-to-supernova-remnant transition using hydrodynamic modeling.
1-D SN models were used as input to study two core-collapse remnants, 
SN 1987A \citep{orlando15} and Cas A \citep{orlando16}.  Full evolution of 3D Type Ia supernova models to SNR stage is relatively new. \cite{orlando20} presented a full 3D MHD simulation of SN 1987A, and \cite{orlando21} performed a similar calculation aimed at
Cassiopeia A. \cite{ferrand19} and \cite{ferrand21} evolved some of the models of \cite{seitenzahl13} to an age of 500 years, focusing on the morphology of the forward and reverse shocks, and the contact discontinuity.  Tycho's SNR appears to have been a fairly normal SN Ia \citep{krause08}, but symmetric initial conditions apparently cannot account for some asymmetry as deduced from spatial power spectra \citep{ferrand19}.  \cite{ferrand21} evolved a quite asymmetric model also from \cite{seitenzahl13} (model N5ddt; see below) again focusing on shock morphology.  
Here we describe a similar evolutionary calculation, to explain a highly unusual young supernova remnant, and requiring the most asymmetric of all the models of \cite{seitenzahl13}.  We consider both morphological and abundance structure, as the target of our study is remarkable in both dimensions.  

\section{A new test case: G1.9+0.3}
\label{obs}

\begin{figure}
\vspace{0.1truein}
\plotone{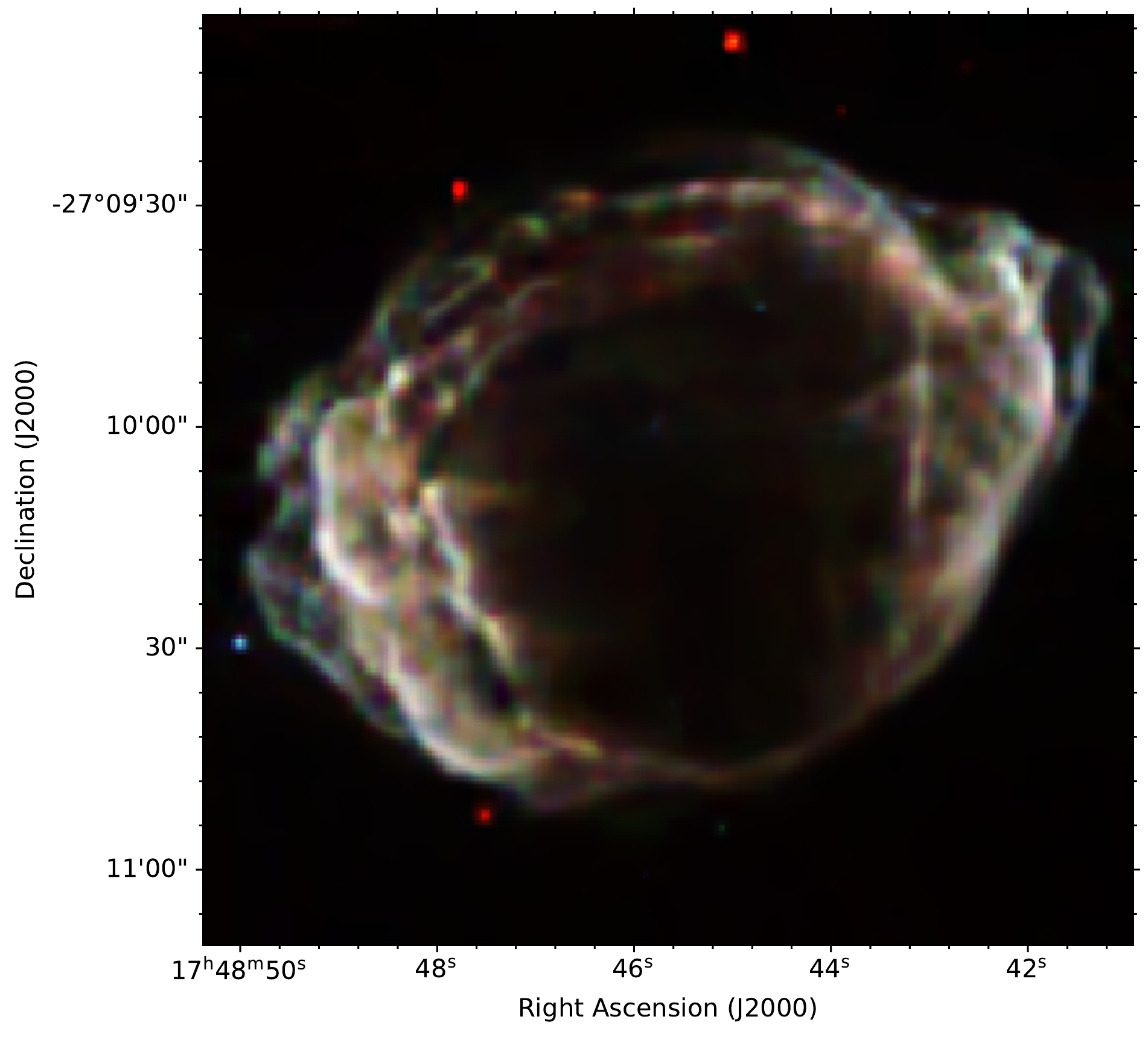}
\caption{G1.9+0.3 in X-rays in 2011 (based on the deep Chandra observations 
described by \cite{borkowski13b}).
Red, 1 -- 3 keV; green, 3 -- 4.5 keV; blue, 4.5 -- 7.5 keV.
\label{xim}}
\end{figure}

One major difficulty afflicts the study of SNRs to learn about their
supernovae: separating ejected material from swept-up surrounding unmodified interstellar medium (ISM)
or modified CSM.  Abundance clues are powerful, but have limitations.  Ideally
one would observe the youngest possible remnant, but large enough for
adequate spatial resolution.  That remnant appears to be the youngest 
Galactic SNR, G1.9+0.3 \citep[][see   Fig.~\ref{xim}]{reynolds08a}.  This object is about $100''$ in diameter, the smallest angular size of any confirmed Galactic SNR. Unfortunately, it is very highly absorbed, with an X-ray column
density of about $5 \times 10^{22}$ cm$^{-2}$ \citep{reynolds09},
implying $A_V \sim 23^m$, so radio and X-rays are the only useful
observational channels.  The angular expansion rate of 0.64 arcsec
yr$^{-1}$ obtained from comparing X-ray images from 2007 and 2009
\citep{carlton11} gives an upper limit for the age of about 160 yr,
less if (as is almost certainly the case) deceleration has occurred;
spatial variations in expansion rate \citep{borkowski14} are
consistent with an age of about 100 yr, or a date of around 1900.
The high extinction would have rendered it unobservable in optical
telescopes of that era.  Furthermore, its X-ray
spectrum is almost entirely synchrotron emission, making it a member
of the small class of X-ray-synchrotron-dominated SNRs.  However, long
observations with {\sl Chandra} have allowed the detection of thermal
emission from small regions \citep{borkowski11a,borkowski13b}, with
spectroscopic widths $\sim 14,000$ km s$^{-1}$ confirming the large
expansion proper motion, refined with a second {\sl Chandra}
observation \citep{carlton11}.  The distance is still uncertain; the
high column, higher than the entire Galactic column along nearby
sight-lines, suggests an association with the Galactic Center, and a
provisional distance of order 8.5 kpc has been assumed.  Nearer would
be very unlikely in view of the high absorption, but too much farther
would make the expansion proper motion unreasonably large.  An H
I observation with the Giant Metrewave Radio Telescope \citep{roy14} has been used to set a lower
limit of 10 kpc, certainly consistent with the known properties of
\src.

While the SN type of \src\ is not absolutely confirmed as Ia, most
indicators point in that direction.  The very high expansion speeds
even after $\sim 100$ yr; the clear presence of supersolar iron (Fe
K$\alpha$) in several regions \citep{borkowski13b}; the absence of any
pulsar-wind nebula in the center (although a neutron star itself would
be too highly absorbed to be detected); and the bilateral symmetry of
the synchrotron X-rays, suggesting a parallel with SN 1006, all point
to a Ia origin.  A core-collapse (CC) scenario can be constructed with
difficulty, but it requires a highly unusual event, with a low mass
but very high energy at explosion, and a low-density surrounding
medium.  We shall assume for the remainder of this work that
\src\ originated in a thermonuclear event.

Careful analysis of the total current {\sl Chandra} exposure of about
1.7 Ms has shown clear evidence for Si and S K$\alpha$ emission, 
as well as Fe K${\alpha}$ \citep{borkowski11a,
  borkowski13b}.  Remarkably, Fe is found at quite large radii, and
the Si/Fe ratio varies considerably.  There is tentative evidence for a broad feature centered at 4.1 keV from the remnant center, attributed to an electron capture in $^{44}$Sc, formed in the decay chain of $^{44}$Ti
and indicating a $^{44}$Ti mass of about $1 \times 10^{-5}\ M_\odot$
\citep{borkowski10,borkowski13a}.  The Fe lines are quite broad,
consistent with the presence of Fe at large projected radii from the
remnant center; \cite{borkowski13b} find FWHM values of 15,000 km
s$^{-1}$.

\begin{figure}
  \vspace{0.1truein}
  \plotone{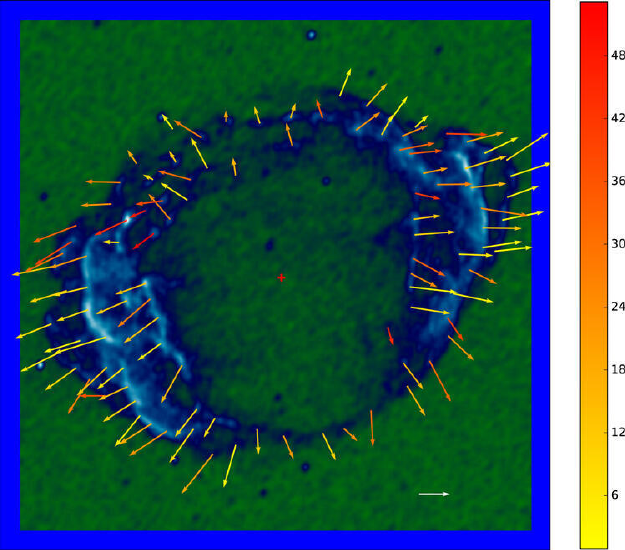}
  \caption
      {Proper motion vectors between 2011 and 2015 \citep{borkowski17},
        color-coded according to deviations from radial directions, as
        measured with respect to the geometrical center (indicated
        with the red X).  The vertical scale shows angles in degrees.
        The white arrow represents an expansion rate of $0.25''$ yr$^{-1}$.}
      \label{expan17}
      \end{figure}

The high observed velocities and proper motions of \src\ indicate that
it is at quite an early evolutionary stage.  Defining the deceleration
parameter $m$ of a region at radius $r$ by $r \propto t^m$,
\cite{borkowski14} compared the 2007, 2009, and 2011 {\sl Chandra}
observations to obtain $m$ values varying spatially, from $\lapprox
0.5$ for the outer blast wave to $\lapprox 0.7$ for inner features
identified with the reverse shock.  In the Sedov self-similar phase,
we have $m = 0.4$ throughout, so the expansion is considerably less
decelerated.  However, the expansion velocities in \src\ vary by a
large factor. \cite{borkowski17} used a nonparametric method to
measure the expansion throughout the remnant and found velocities
varying strongly with position, over a factor of 5 in magnitude
(Fig.~\ref{expan17}).  In particular, expansion to the north was found
to be several times slower than in other directions, strongly
suggesting a recent encounter with denser material.  Such an encounter
could explain the peak in radio intensity found there, as a higher
density of accelerated electrons would be expected, but the slower
shock velocity would result in a lower maximum accelerated electron
energy, explaining the lower X-ray brightness to the north.  Such
denser material almost certainly originated in asymmetric pre-SN mass
loss rather than in the dynamics of the explosion itself, although
such a possibility cannot be ruled out.

The total mass of shocked ejecta in \src\ is still quite small.  An
analysis based on the average expansion rate found 0.033 $M_\odot$
\citep{carlton11}, while one based on scaling an energetic explosion
model WS15DD3 \citep{hachinger13} produced about twice as high a
value, 0.064 $M_\odot$ \citep{borkowski13b}.  While these values are
much less than the total ejecta mass, we observe Fe to make up a
significant fraction of material already shocked.

The presence of high-velocity shocked Fe is especially problematic in
\src\, as it was presumably near the outer edge of the expanding
ejecta at a very early time.  One might expect evidence of such fast
Fe in observations of Type Ia supernovae themselves; while
high-velocity features are seen in a large fraction of all SNe Ia
\citep[e.g.,][] {maguire14}, Fe is rarely seen, and when it is, the
estimated mass is very small.  Fitting of SN spectra with stratified
models \citep[``supernova tomography," e.g.,][]
{stehle05, tanaka08, sasdelli14, mazzali15} also show little evidence
for high-velocity Fe. This raises the question of whether
current Type Ia supernova models will evolve to resemble \src\, in
this property as well as in the bulk expansion velocity and observed
spatial heterogeneity of Si, S, and Fe.  In this paper, we address
this question, using two multi-dimensional SN Ia models as initial
conditions and following their expansion into uniform media with the
VH-1 hydrodynamics code.

\section{Simulations}

The VH-1 hydrocode is a conservative, finite-volume code evolving the
Euler equations for an ideal gas, based on the Piecewise Parabolic
Method with the Lagrange-Remap implementation \citep{colella84}.  It
is particularly well suited to resolving shocks and other
discontinuities, and has been extensively used for the study of
supernova remnants \citep[see, for instance, 3D hydrodynamic
simulations of Type Ia SNRs:][]{warren13}. It can be used
in one, two, or three dimensions; in two dimensions, either with $r,
z$ cylindrical coordinates, or $r, \theta$ spherical coordinates, and
in three dimensions, using the overset spherical grids known as
``Yin-Yang'' \citep{kageyama04} that eliminate the singularities on
the axis of symmetry.  Here, we shall perform 2D simulations in
cylindrical $r, z$ coordinates, and 3D ones using the Yin-Yang grids.  
We do not include effects of energy losses due to cosmic rays, which
can increase the compressibility of the gas \citep{blondin01}, 
but use a uniform adiabatic index of $5/3$.  Our 2D cylindrical grid has a resolution of $512 \times 1024$, and expands to follow the evolution of the SNR. The 3D simulation also expands with the blast wave. It has an angular resolution of $0.24^\circ$ (384 $\times$ 1152 angular zones in each of the Yin and Yang grids).  For the radial
coordinate, we cover the inner half of the grid with 84 zones and
the outer half with 300, to maximize resolution in the region between
the shocks.  This gives a maximum fractional resolution of
$1.65 \times 10^{-3}$ of the maximum grid radius for the outer 50\% of that maximum radius.

We use two initial Type Ia supernova models (F.~R\"opke, private
communication).  Table~\ref{compos} gives the compositions of both. 
Our 2D model comes from the family described in
\cite{kasen09}, in which 2000 otherwise identical 1.4 $M_\odot$ C/O
white dwarfs (50\% each) were ignited with varying numbers and
locations of ignition points. The model was provided with a 
resolution of $512 \times 1024$. Our 3D model is model N3 from 
\citet{seitenzahl13}, with
three slightly off-center ignition points.  The mass is 1.40
$M_\odot$, the radius $1.96 \times 10^8$ cm, and the central density
$\rho_c = 2.9 \times 10^9$ g cm$^{-3}$.  The composition is also equal
parts $^{12}$C and $^{16}$O (with 2.5\% $^{22}$Ne to account for the
approximately solar metallicity of the ZAMS progenitor). The white dwarf undergoes a delayed
detonation; a detailed discussion of the ignition setup is given in
\citet{seitenzahl13}.  This model was chosen as the most asymmetric of
the suite of 14 models in \citet{seitenzahl13}.  We describe each
simulation in more detail below.

\subsection{2D Simulation}

\begin{figure}
\centerline{\includegraphics[scale=0.6]{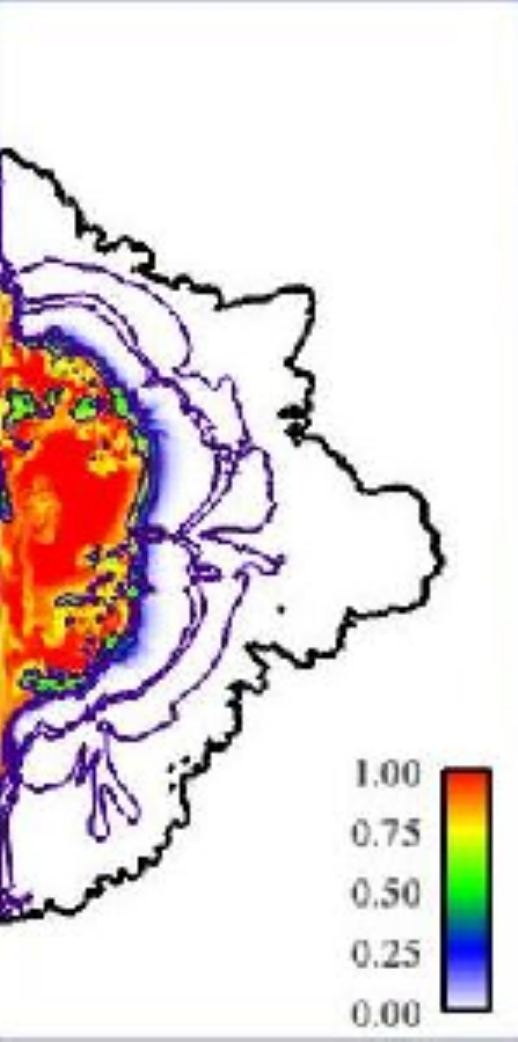} 
\hskip0.5truein \includegraphics[scale=0.7]{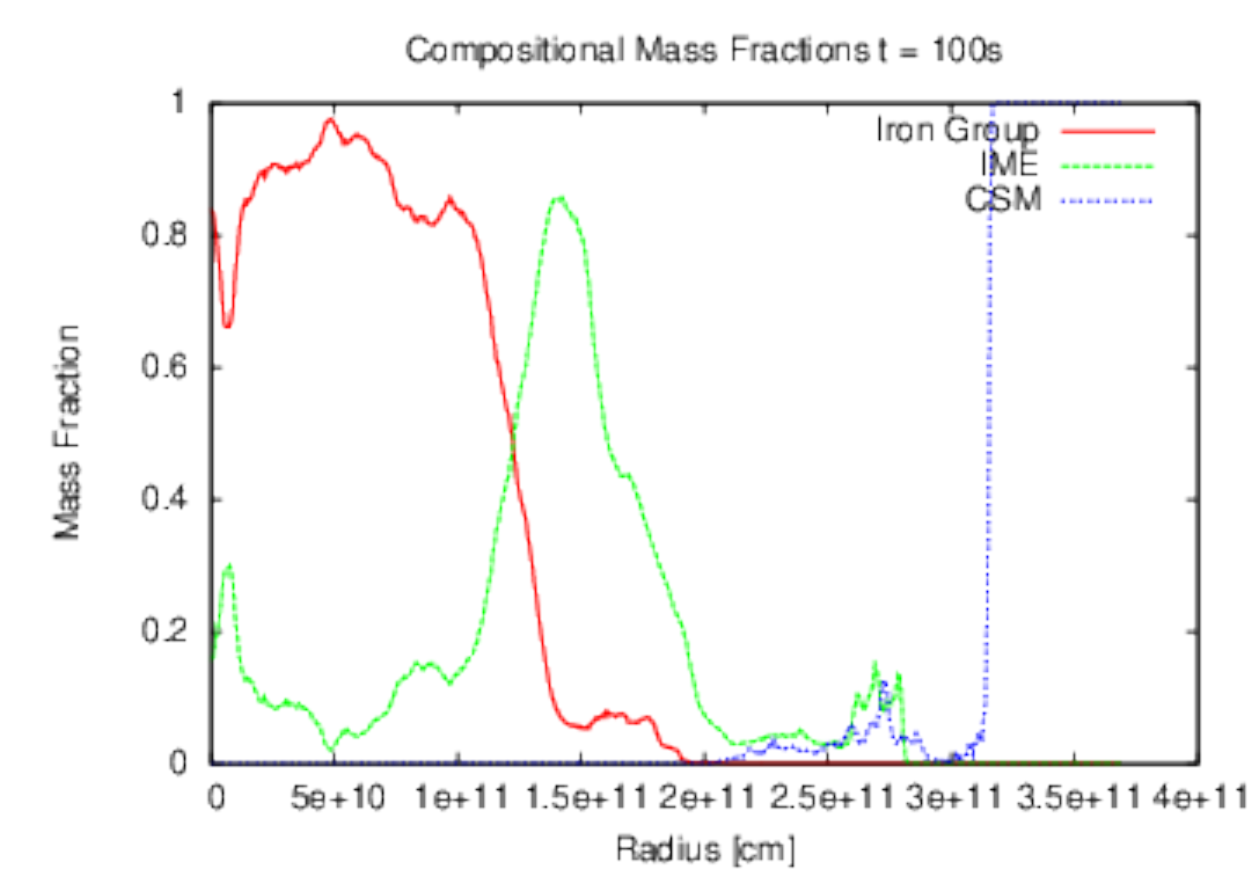}}
\caption{2D model. Left: 2-D distribution.  Black contour: contact
  discontinuity between CSM and ejecta.  Purple contour: IME's.  Color
  scale: Mass fraction of iron-group elements.  Right:  Initial mass
fractions, azimuthally averaged.}
\label{2Ddist0}
\end{figure}

\begin{figure}
\centerline{\includegraphics[scale=0.6]{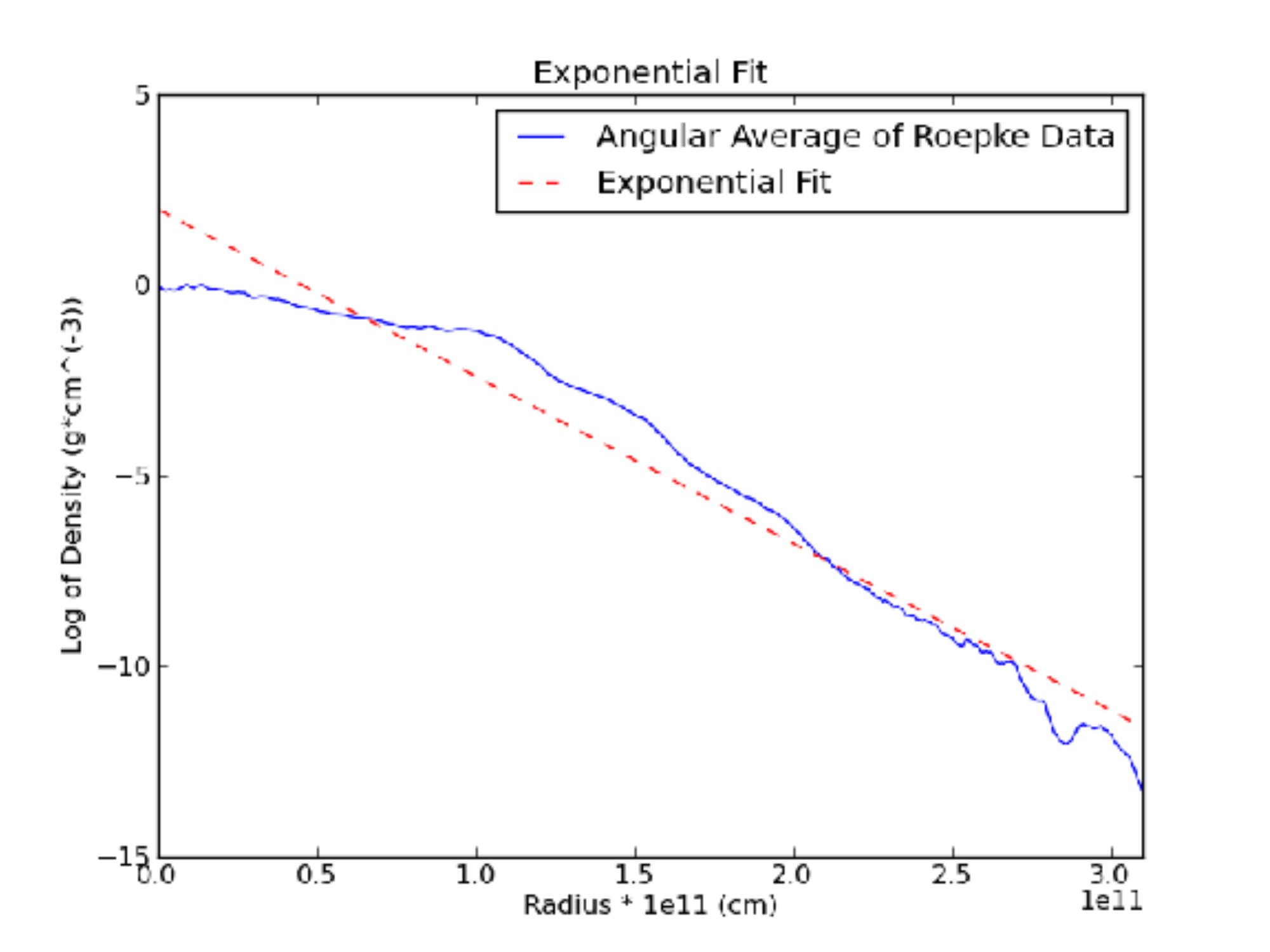}}
\caption{Azimuthally averaged density of 2D model, with superimposed
least-squares fit to an exponential, which describes the data fairly well.}
\label{expofit2d}
\end{figure}

Our 2D simulation was conducted at the same resolution as that of
the initial model, $(r, z) = (512, 1024)$.  
The initial model at an age of 100 s is shown in Figure~\ref{2Ddist0},
showing the location and iron content of ejecta, as well as the
location of the contact discontinuity.  The composition is given
in Table~\ref{compos}; the total asymptotic kinetic energy is 
$1.45 \times 10^{51}$ erg, and the total mass 1.408 $M_\odot$. 
The model initially expands
into an unrealistic number density of $10^6$ cm$^{-3}$, but it is
straightforward to scale this to an arbitrary value.  (Figure~\ref{expofit2d} 
shows that the outermost ejecta density is more than three orders of 
magnitude larger; the ram pressure $\rho v_{\rm sh}^2$ is about 
$10^5$ dyn, for an initial expansion velocity of $10,000$ km s$^{-1}$, 
while the unrealistically high-density initial CSM still has a pressure 
of only $nkT \sim 10^{-4}$ dyn, even for a temperature of $10^6$ K.)  
The radial distribution of elements is also shown in Figure~\ref{2Ddist0}, 
where we combine elements into intermediate-mass elements (IME) and
iron-group elements (IGE).  CSM represents all surrounding
material. The azimuthally
averaged density profile is fairly well fit by an exponential
\citep{dwarkadas98}, as shown in Figure~\ref{expofit2d}.

\begin{deluxetable}{lcc}
\tablecolumns{2}
\label{compos}
\tablecaption{Initial Model Compositions}
\tablehead{
\colhead{Element} &2D Model Mass ($M_\odot$) & 3D Model Mass ($M_\odot$)}
\startdata
C+O         & 0.0853 & 0.0624\\
IME         & 0.378  & 0.121\\
Iron-group  & 0.945  & 1.22\\
$^{56}$Ni   & 0.698  & 1.10\\ 
Total       & 1.41  & 1.40\\
\enddata
\end{deluxetable}

\begin{figure}
\centerline{\includegraphics[scale=0.6]{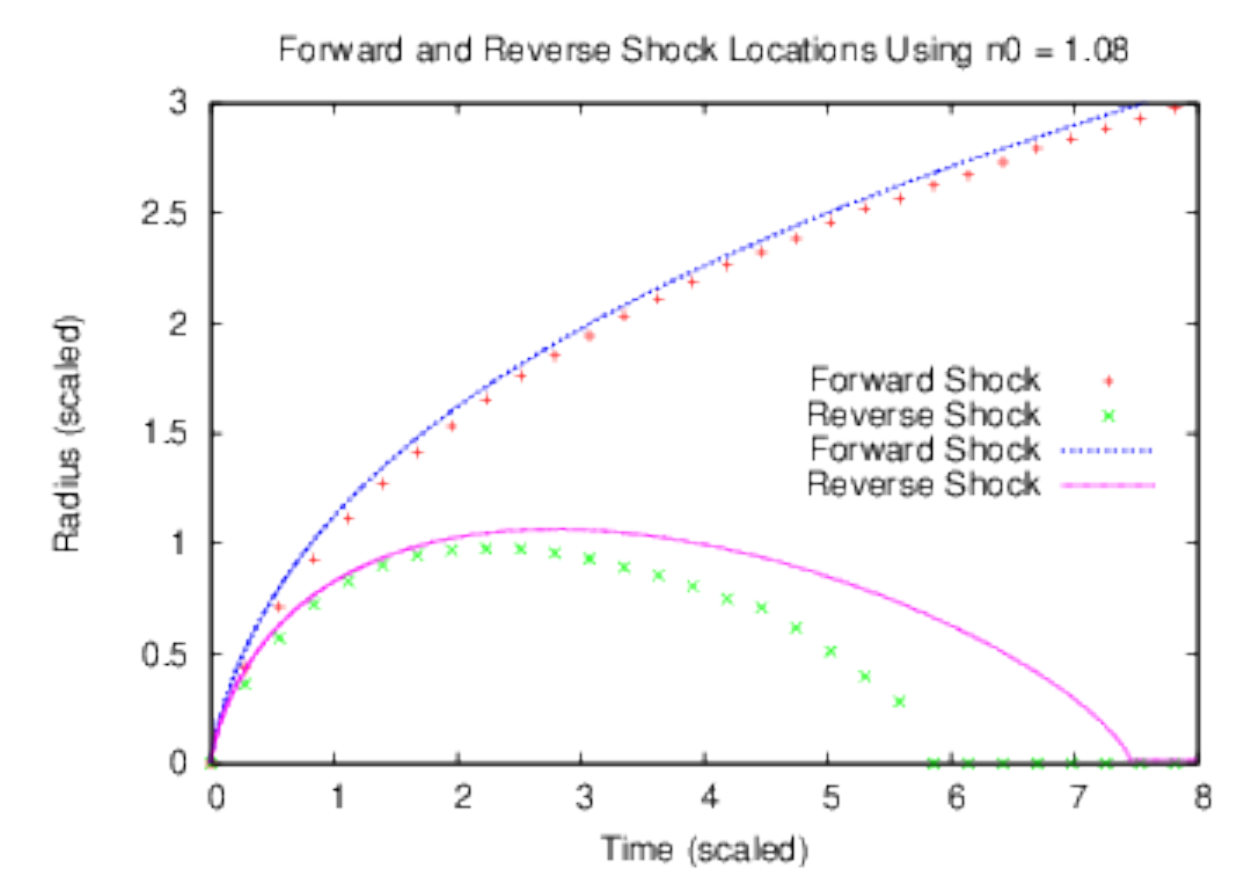}}
\caption{Forward and
reverse shock locations.  Points:  Azimuthal averages from 2D simulation.
Curves: Results from a 1D simulation using the exponential fit to the
initial density profile (Fig.~\ref{expofit2d}).  The time and radius are
scaled to the present age of G1.9+0.3.  At these early times, the
shock positions in the exponential model are quite close to those
from the 2D calculation.}
\label{shockradii}
\end{figure}

The model was then evolved for two different values of constant
upstream density, one chosen to match the value of 0.022 cm$^{-3}$
found from the expansion measurements of \cite{carlton11}, and the other
a value of 1.08 cm$^{-3}$, calculated to bring the reverse shock
farther in (see below).  Figure~\ref{shockradii} shows the
angle-averaged forward and reverse shock locations for the
high-density run, compared to a 1D simulation using the exponential
fit to the initial density profile shown in Fig.~\ref{expofit2d}.

\begin{figure}
\centerline{\includegraphics[scale=0.7]{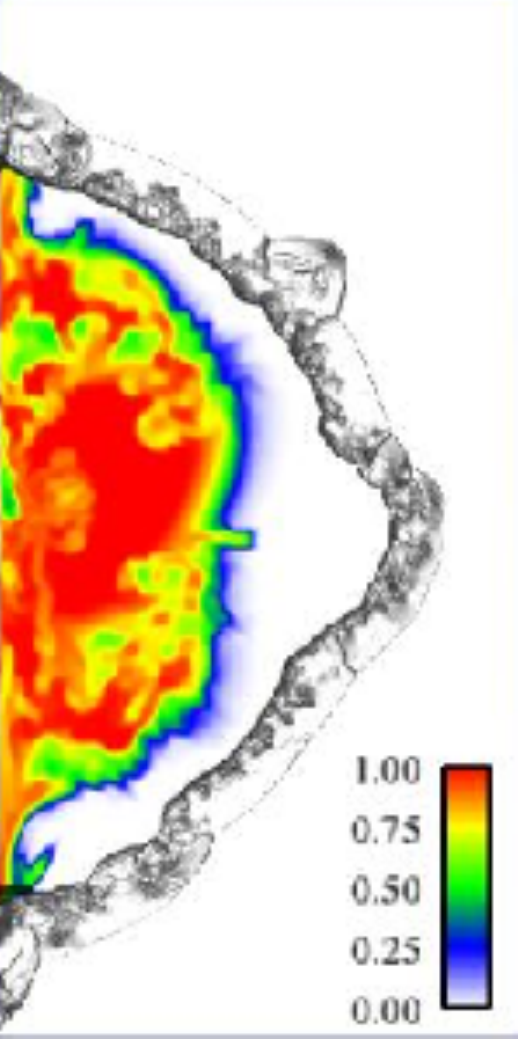}\hskip0.4truein
  \includegraphics[scale=0.7]{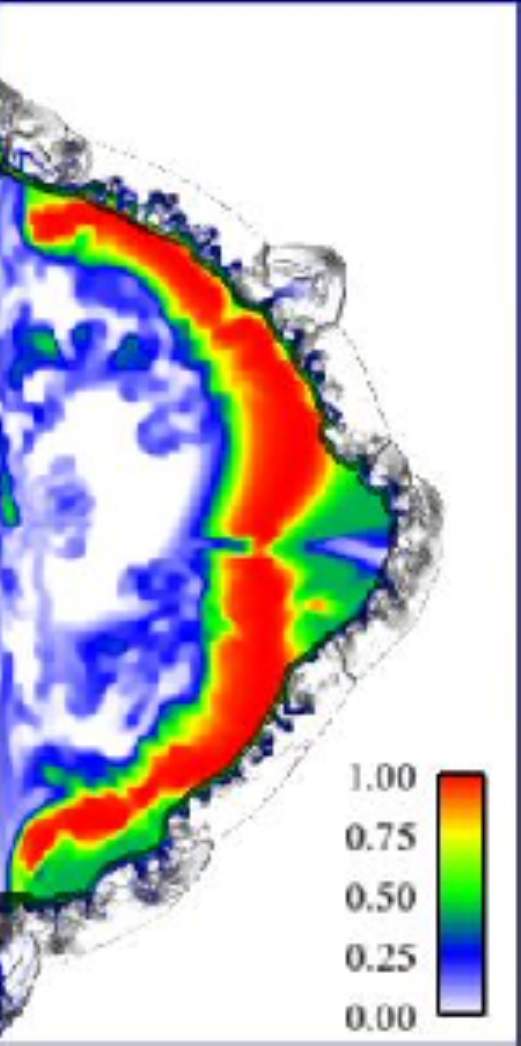}\hskip0.4truein
  \includegraphics[scale=0.7]{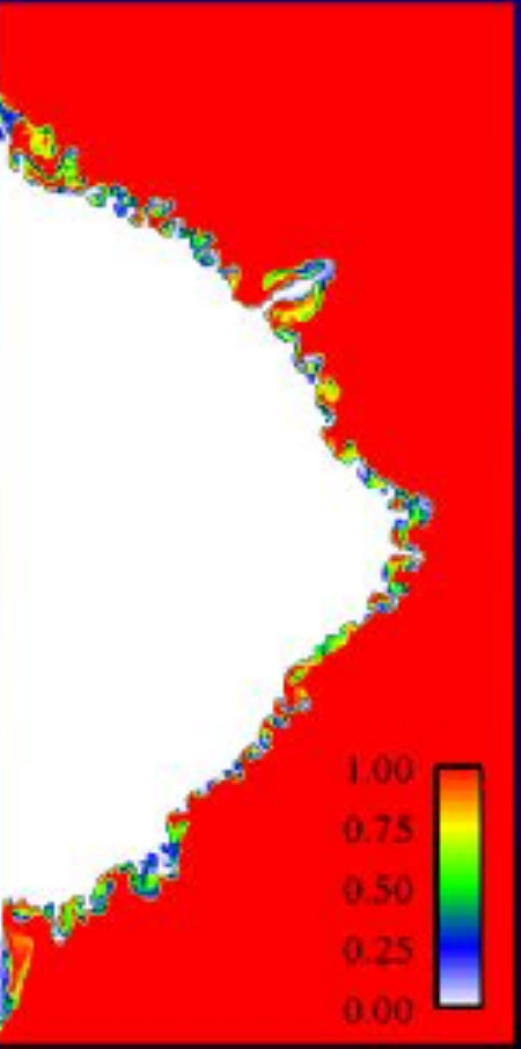}}
\caption{2D model at the size of G1.9+0.3 ($r = 2.0$ pc), for
$n_0 = 0.022$ cm$^{-3}$.  Left:
Iron-group elements.  Center:  Intermediate-mass elements. Right:
Circumstellar material.  All are given as mass fractions.}
\label{2Ddistslo}
\end{figure}

Figure~\ref{2Ddistslo} shows the distribution of elements at the current
size of G1.9+0.3 ($r = 2.0$ pc), as mass fractions, in the low-density
calculation.  The contact discontinuity is visible as the bound of 
IME's; iron-group elements are well inside at all radii (excepting an
artifact along the polar axis).  The relatively low resolution of the
initial 2D model results in a slight shape distortion in the outermost
contour visible in Fig.~\ref{2Ddist0}.  In the low-density case, this slightly
rectangular shape persists to the age of G1.9+0.3, but in the more highly evolved
high-density case, it has evolved away by this time.  The remnant age is 104 yr for
the low ambient density; the total swept-up mass is 0.026 $M_\odot$.
About 0.06 $M_\odot$ of ejecta have been shocked, quite close to
one of our estimates based on the observed expansion rate.  However, 
only a minuscule amount, $7 \times 10^{-5}\ M_\odot$, of that is 
iron-group elements -- clearly inconsistent with the observations.

\begin{figure}
\centerline{\includegraphics[scale=0.7]{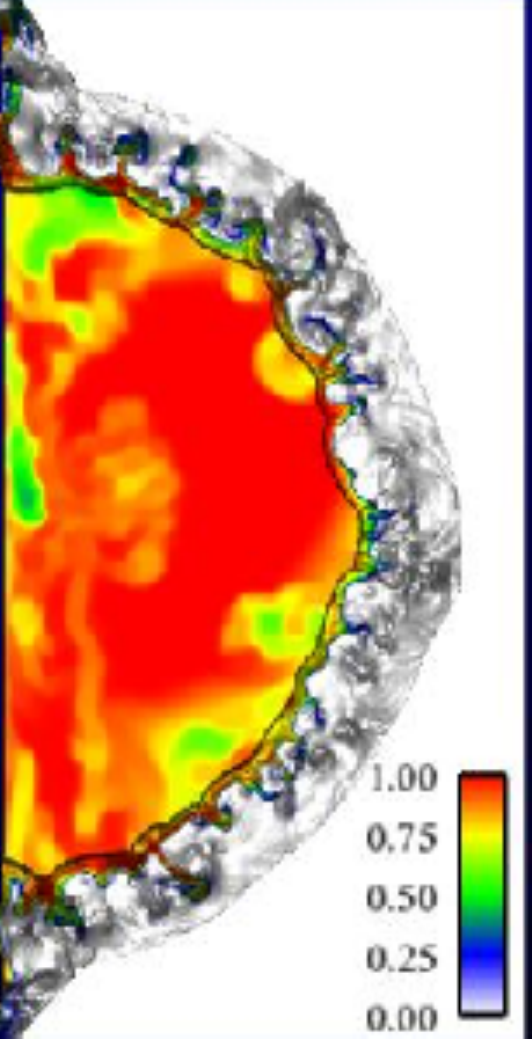}\hskip0.4truein
  \includegraphics[scale=0.7]{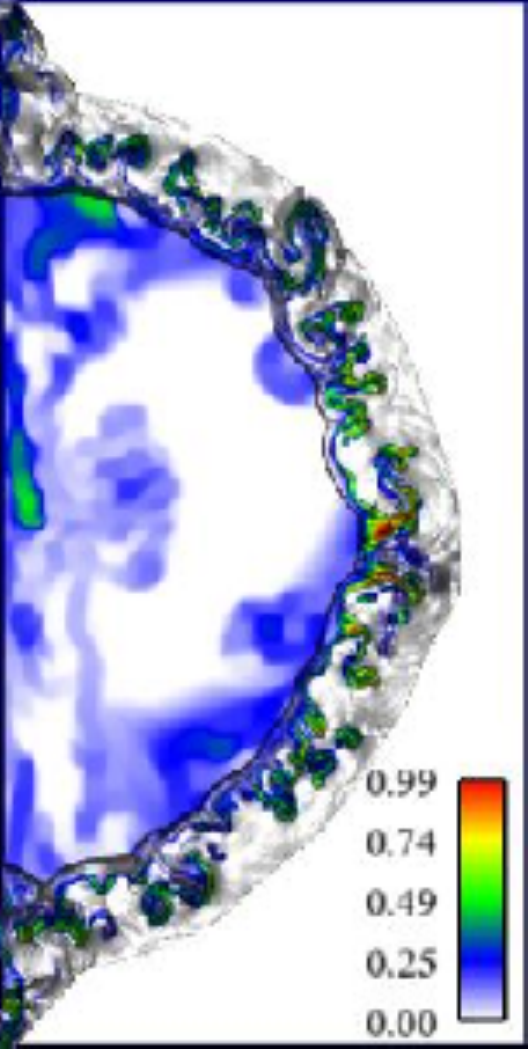}\hskip0.4truein
  \includegraphics[scale=0.7]{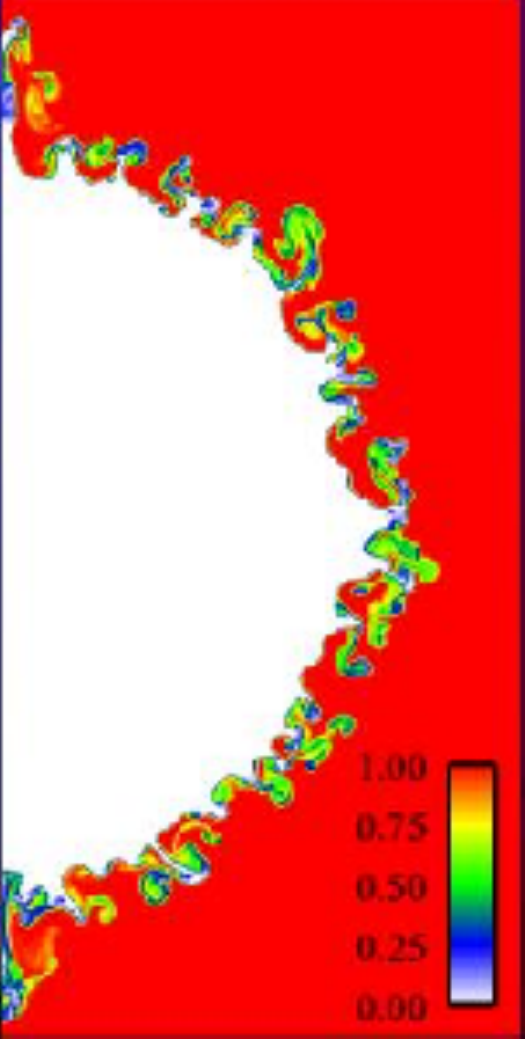}}
\caption{2D model at the size of G1.9+0.3 ($r = 2.0$ pc), for $n_0 =
  1.08$ cm$^{-3}$ (but at a much later age than appropriate for
  G1.9+0.3).  Left: Iron-group elements.  Center: Intermediate-mass
  elements. Right: Circumstellar material.  All are given as mass
  fractions.}
\label{2Ddistshi}
\end{figure}

\begin{figure}
\centerline{\includegraphics[scale=0.6]{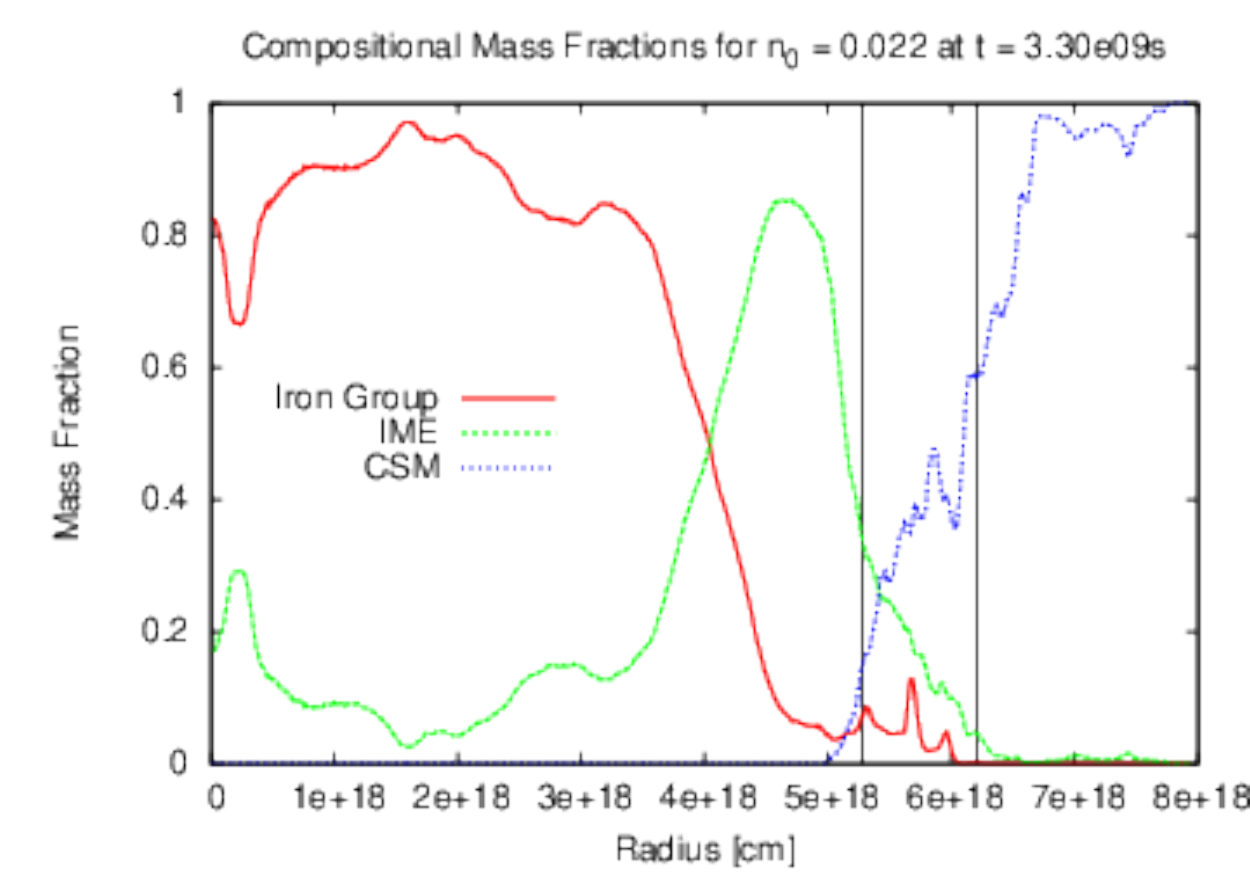} \hskip0.5truein
  \includegraphics[scale=0.6]{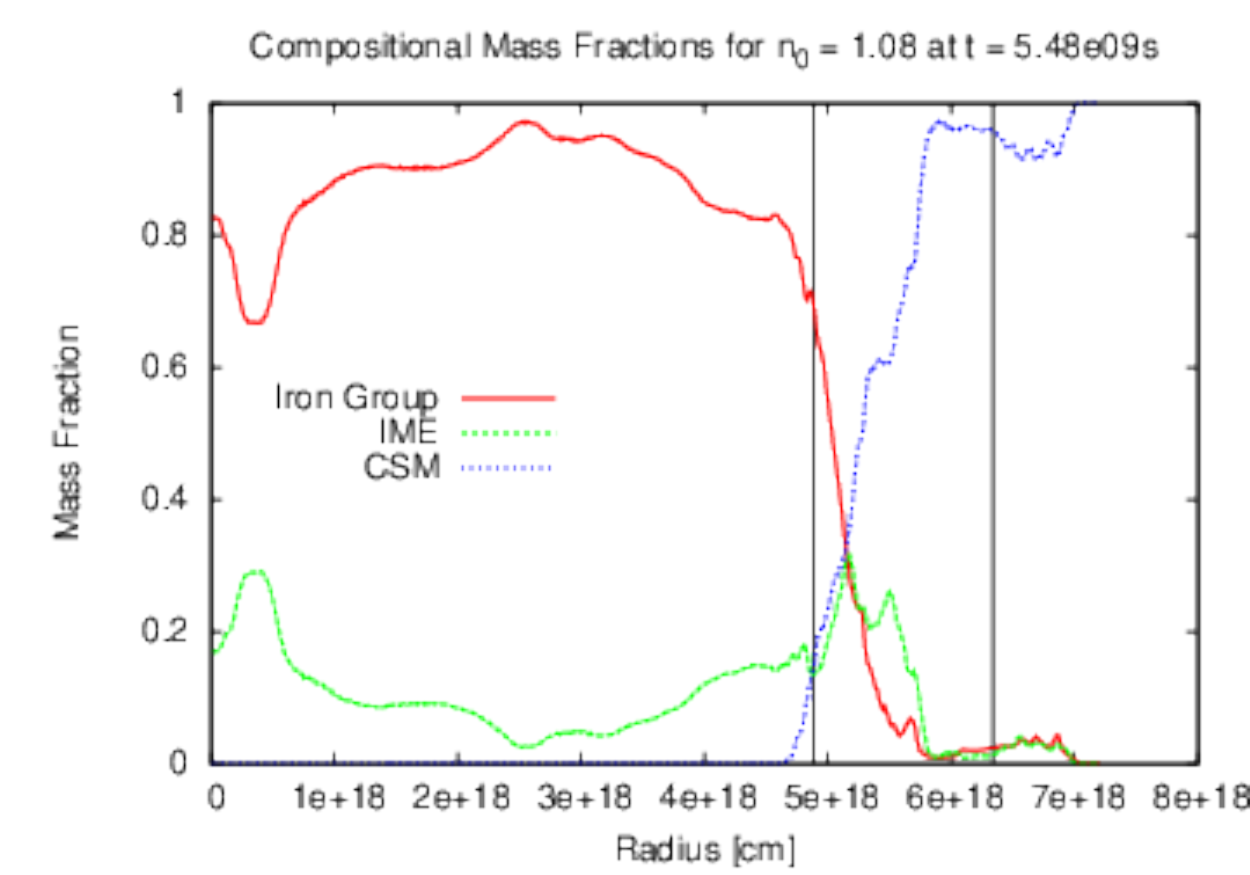}}
\caption{Mass fractions at the size of G1.9+0.3, for the two values of
ambient density in the 2D simulation.  Left:  $n_0 = 0.022$ cm$^{-3}$.  Right:
$n_0 = 1.08$ cm$^{-3}$.  Vertical lines indicate the average positions
of forward and reverse shocks. Only material at larger radii than
the reverse shock would be visible in X-rays.}
\label{2Dmassfracs}
\end{figure}

Figure~\ref{2Ddistshi} shows the elemental distribution for the
high-density model, at an age of 173 yr.  This age is of course inconsistent
with the observational data, and the shock velocity and expansion rate
are far too low, but some such combination of unrealistic parameters is
required to obtain a significant fraction of shocked iron from this
initial model.  Iron-rich material is
nearing the reverse shock.  (Note the much more developed
Rayleigh-Taylor structures beyond the contact discontinuity).
Figure~\ref{2Dmassfracs} compares the azimuthally averaged mass
fractions for the two models.  In the low-density case that matches
the size and expansion rate of G1.9+0.3, very little iron 
($\sim 7 \times 10^{-5}\ M_\odot$) has passed
through the reverse shock to become heated and visible in X-rays.  The
very high density of the other case is required so that at an age of a
few hundred years, a significant (but still small) amount of iron has
been shocked, as the observations require.

\subsection{3D simulation}

\begin{figure}
  \centerline{\includegraphics[width=17truecm]{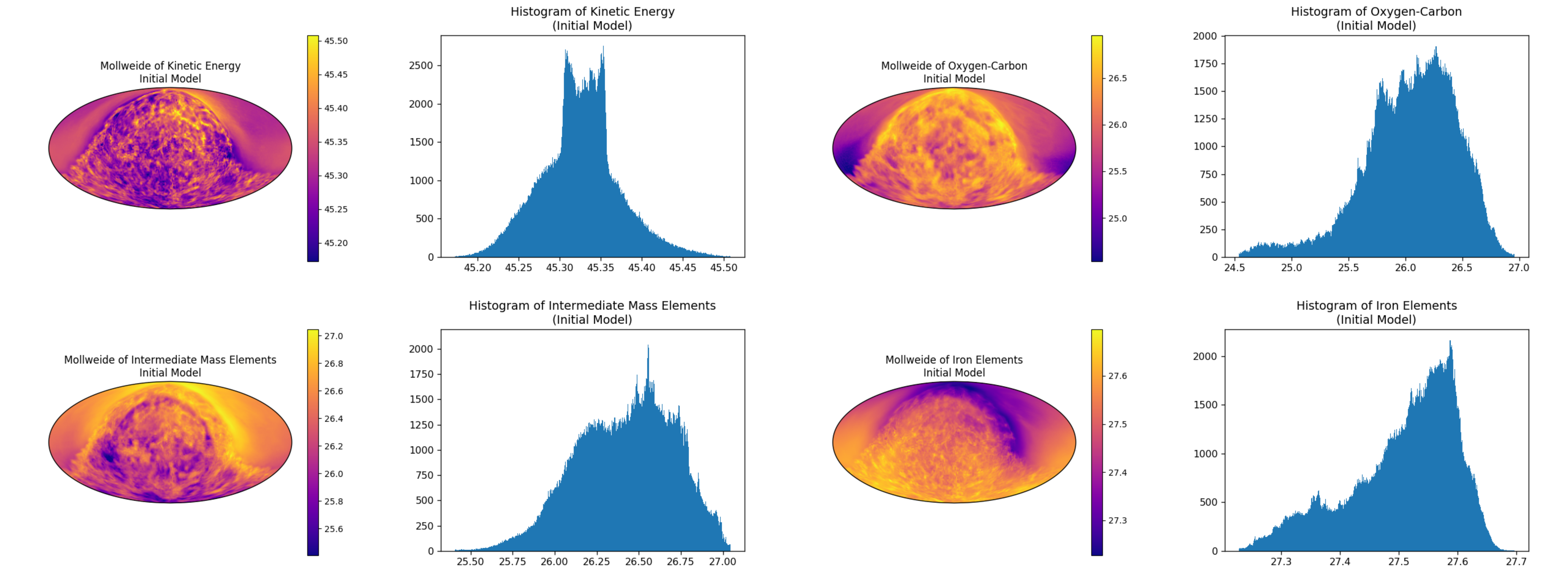}}
  \caption{Mollweide (equal-area) projections of the 3D model N3, for
    (left to right, top to bottom): kinetic energy, and total
    (shocked and unshocked) 
    masses of C/O, IMEs, and IGEs.  Histograms of values
  show the distributions.  Subsequent composition images show shocked
  masses only.}
    \label{initialmodel}
\end{figure}

\begin{figure}
  \centerline{\includegraphics[width=8truecm]{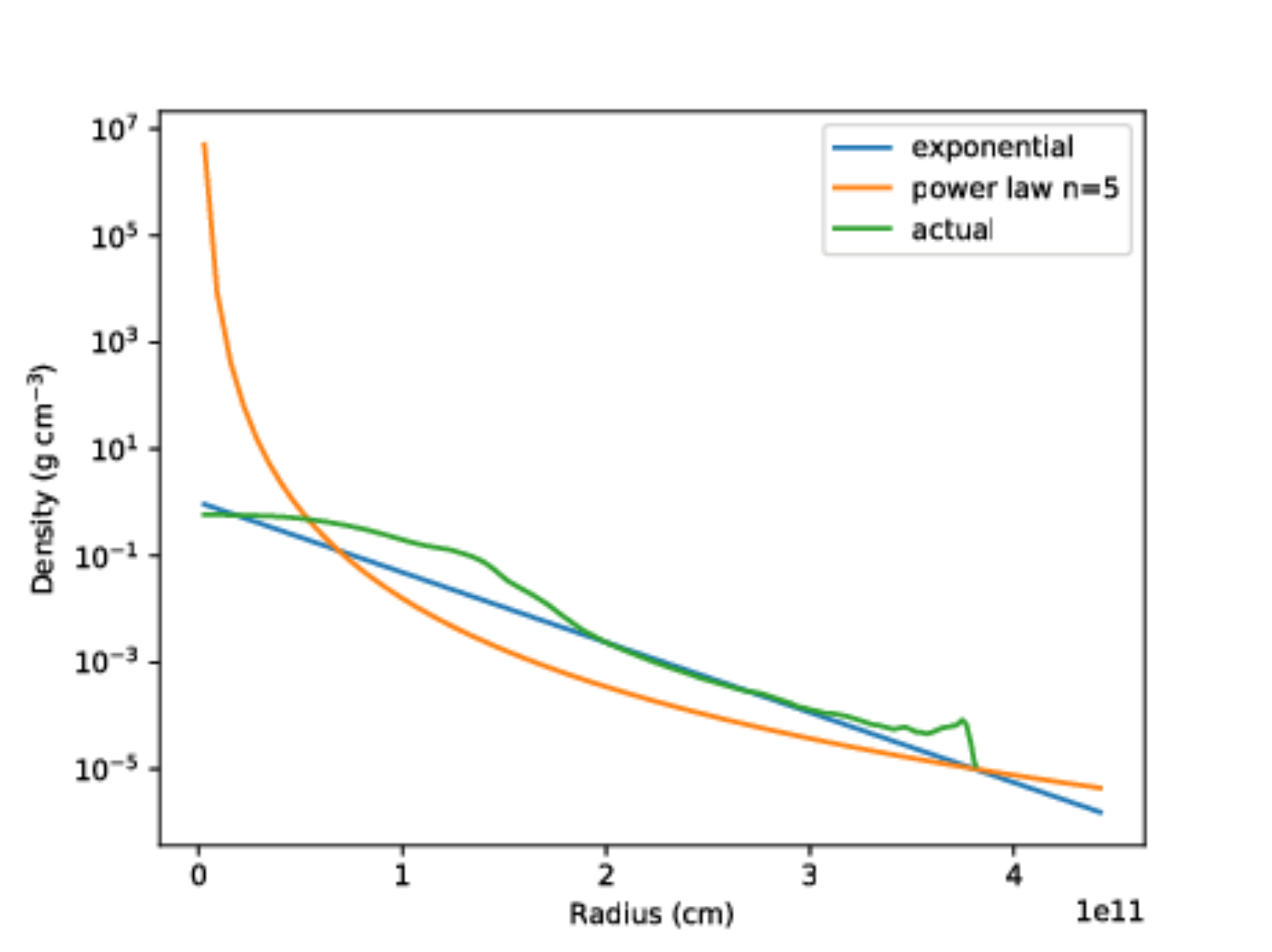}
  \hskip0.5truecm\includegraphics[width=8truecm]{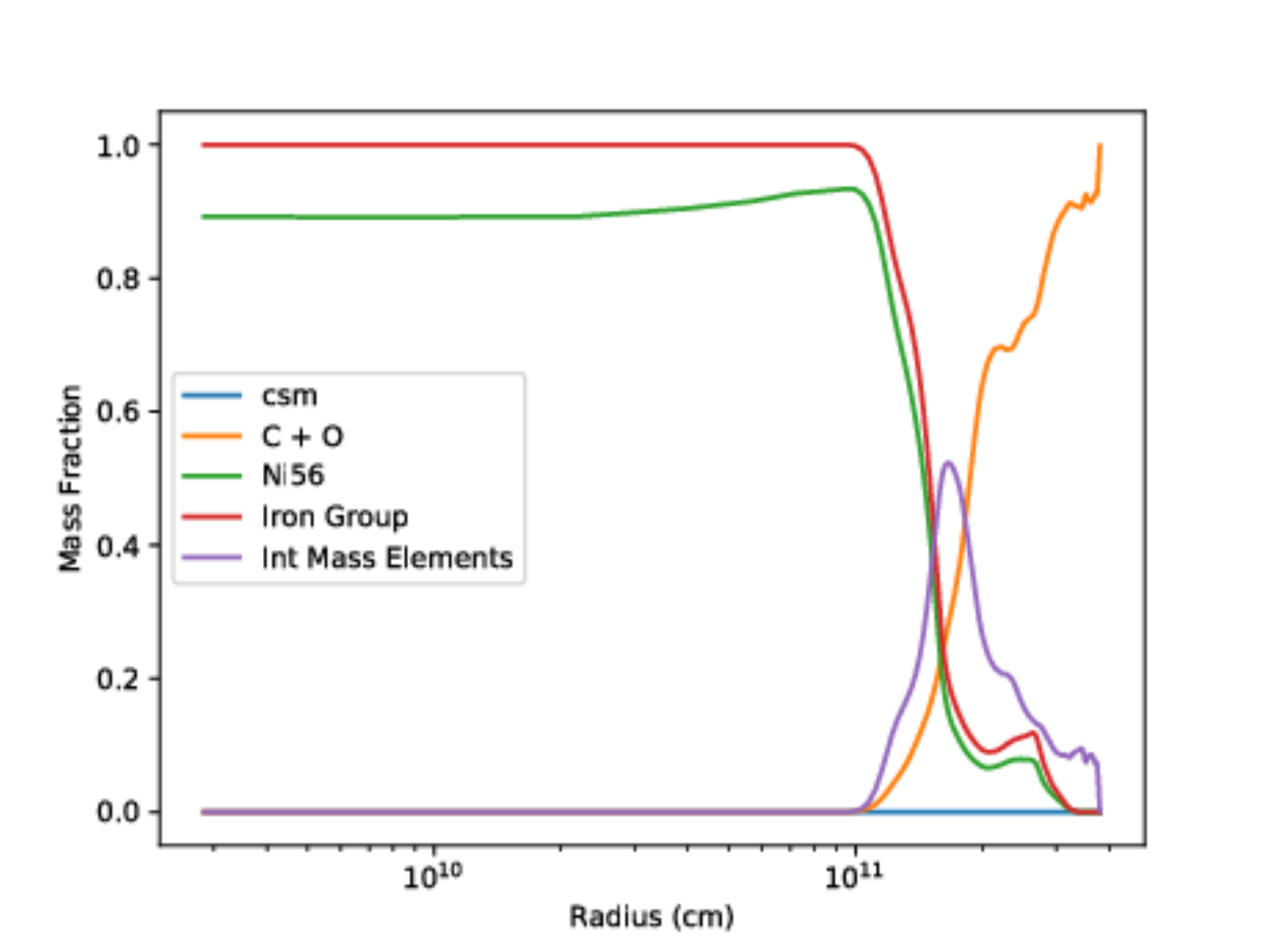}}
  \caption{Left: Angle-averaged density of the initial 3D model.  Also
    shown are exponential and power-law curves for comparison (not
    fits). Scaling radius for exponential:  $3.3 \times 10^{10}$ cm. Power-law index: 5.5. Right: Angle-averaged mass fractions of different species.}
  \label{density}
  \end{figure}

Our 3D model, model N3 from \cite{seitenzahl13}, has a composition given
in Table~\ref{compos}.  The total kinetic energy is $1.61 \times 10^{51}$ erg,
and the total mass 1.40 $M_\odot$.  We illustrate the initial spatial 
distribution of elements and of kinetic energy using Mollweide (equal-area)
projections of the entire surface.  Figure~\ref{initialmodel} shows
the sky distribution of kinetic energy (KE) and ejecta composition
(masses) in three bins (unburned C-O, IMEs, and Fe-group).  Scales 
refer to masses in g pix$^{-1}$, and KE in erg pix$^{-1}$, where one 
pixel subtends $1.72 \times 10^{-5}$ sr.  
Substantial asymmetries are evident.  A ring of enhanced IMEs correlates
with an absence of iron-group elements, which are enhanced in the opposite
hemisphere.  Histograms show the distribution of values in each plot.  
The strong variation in distribution of Fe-group elements reflects the
substantial anisotropy of the initial model.  
Angle-averaged total density and mass fractions are shown in
Fig.~\ref{density}, along with an exponential and power-law (not fit)
for comparison. The exponential shown has an $e$-folding radius
of $3.3 \times 10^{10}$ cm, and the power-law index is 5.5.

\begin{figure}
  \centerline{\includegraphics[width=17truecm]{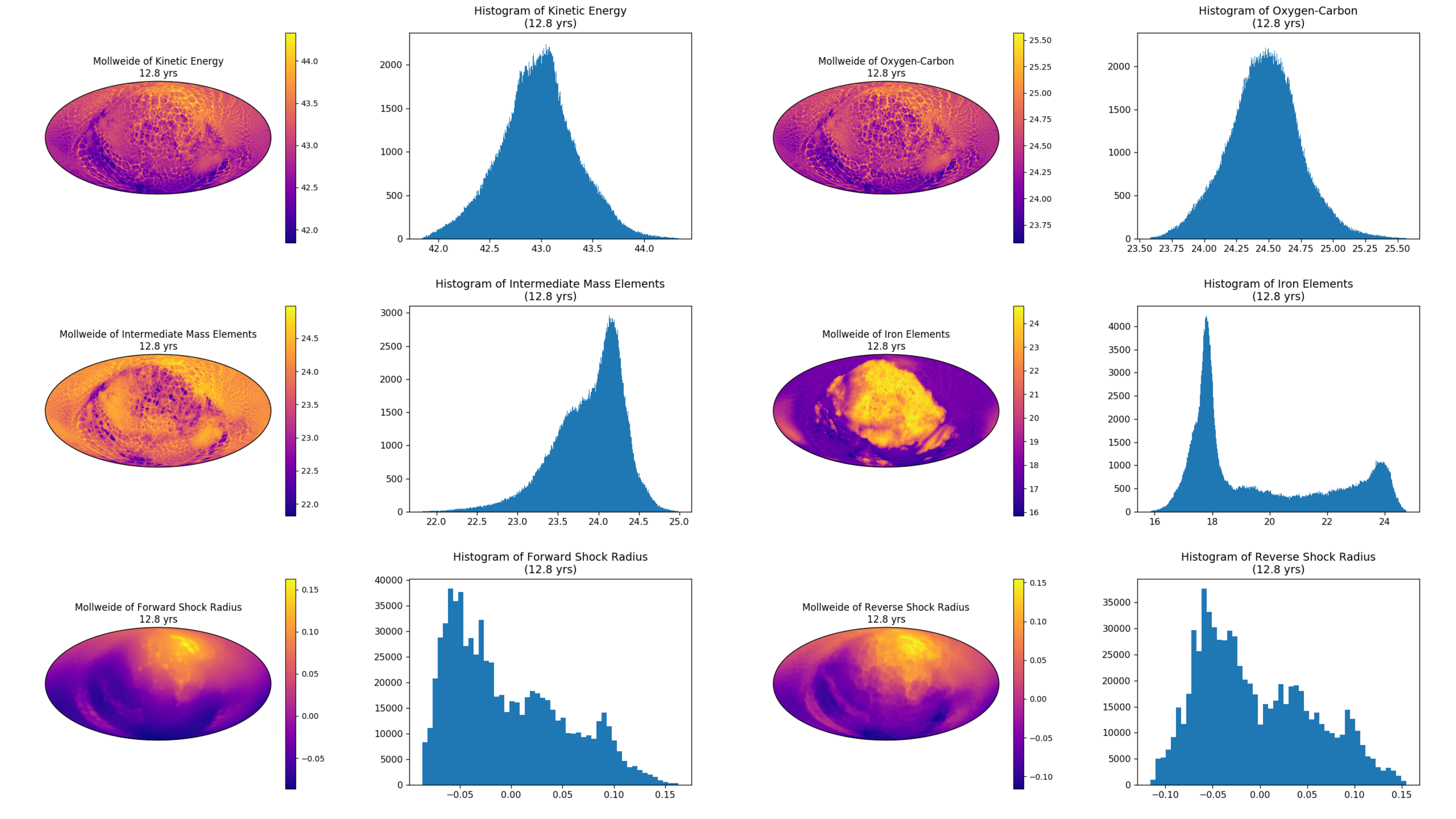}}
  \caption{Mollweide images of (left to right, top to bottom): kinetic energy, 
    shocked CO mass, shocked IMEs, shocked Fe-group elements, and forward
    and reverse-shock radii, all at an age of 13 years. Histograms show the distribution of values in each case.  Energy and masses are in units of log(erg pix$^{-1}$) and log(g pix$^{-1}$), respectively, 
    where one pixel subtends $1.718 \times 10^{-5}$ sr.  Shock radii are given as log($r/r_{\rm
      av}$), where $r_{\rm av}$ is the mean radius. (NOTE: The online article when published will contain animations of the forward and reverse shocks and the three categories of shocked material.)}
    \label{13yr}
\end{figure}

\begin{figure}
  \centerline{\includegraphics[width=17truecm]{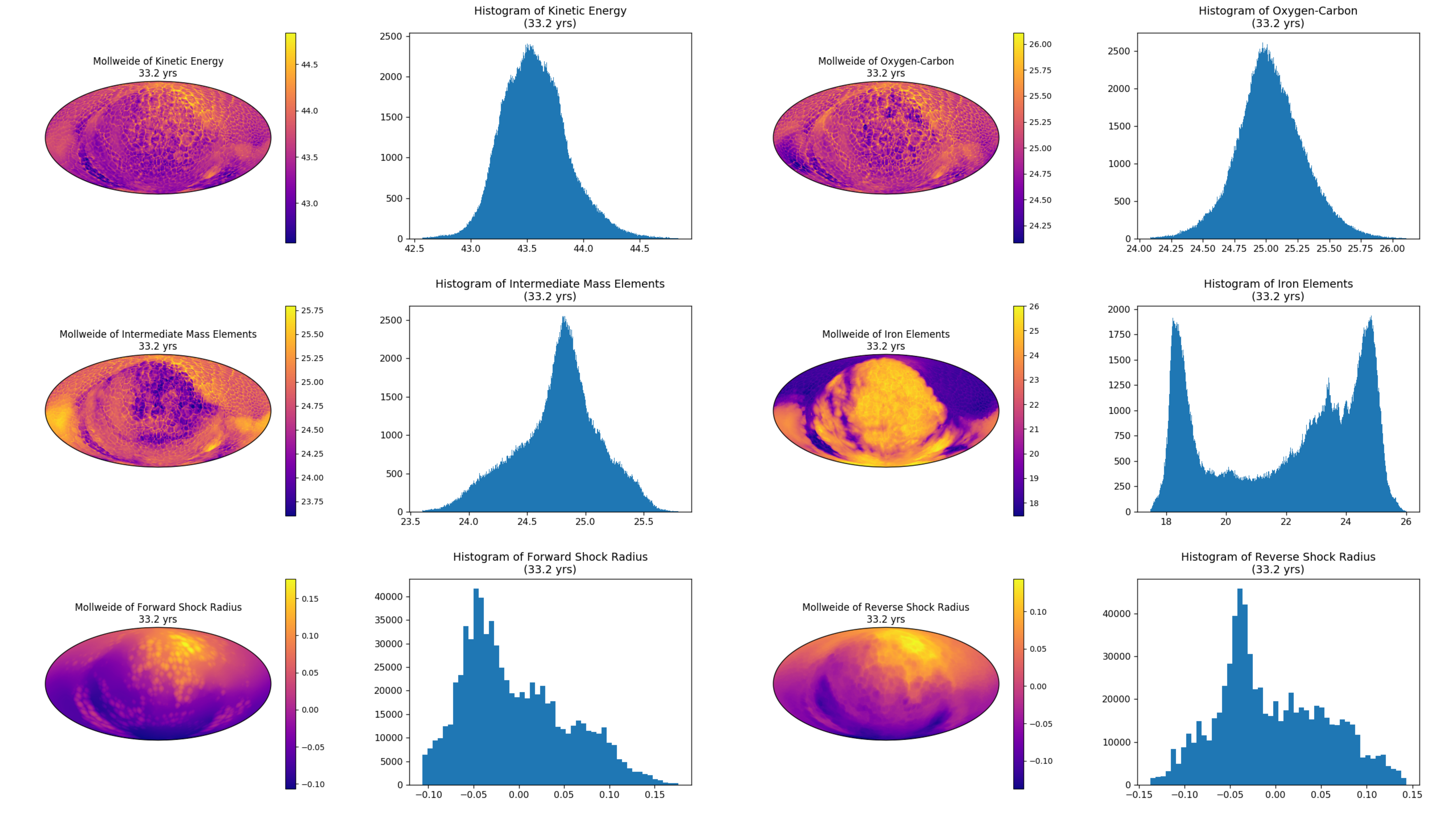}}
  \caption{As in Fig~\ref{13yr}, at an age of 33 years.}
    \label{33yr}
\end{figure}

\begin{figure}
  \centerline{\includegraphics[width=17truecm]{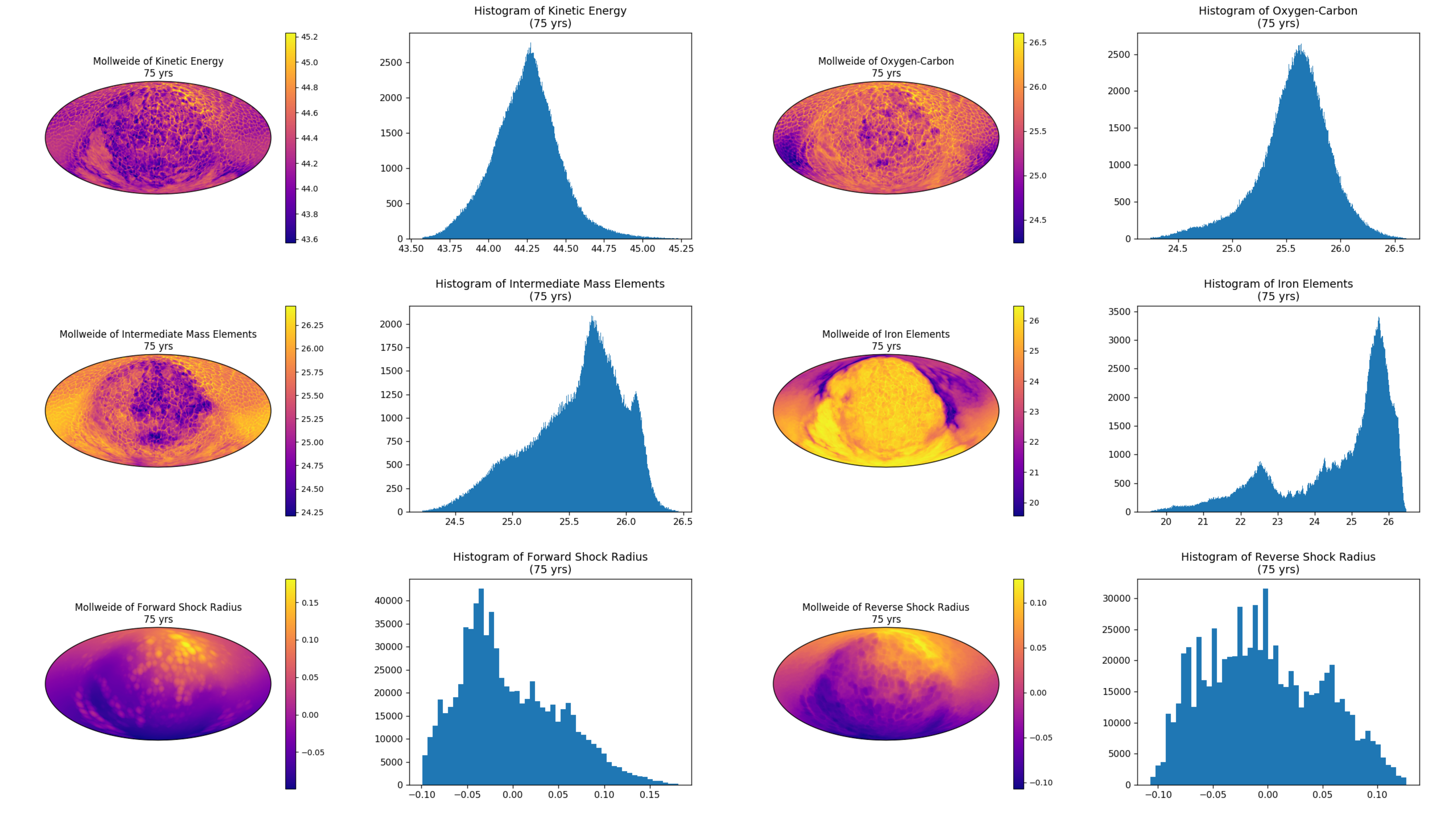}}
  \caption{As in Fig~\ref{13yr}, at an age of 75 years.}
    \label{75yr}
\end{figure}

\begin{figure}
  \centerline{\includegraphics[width=17truecm]{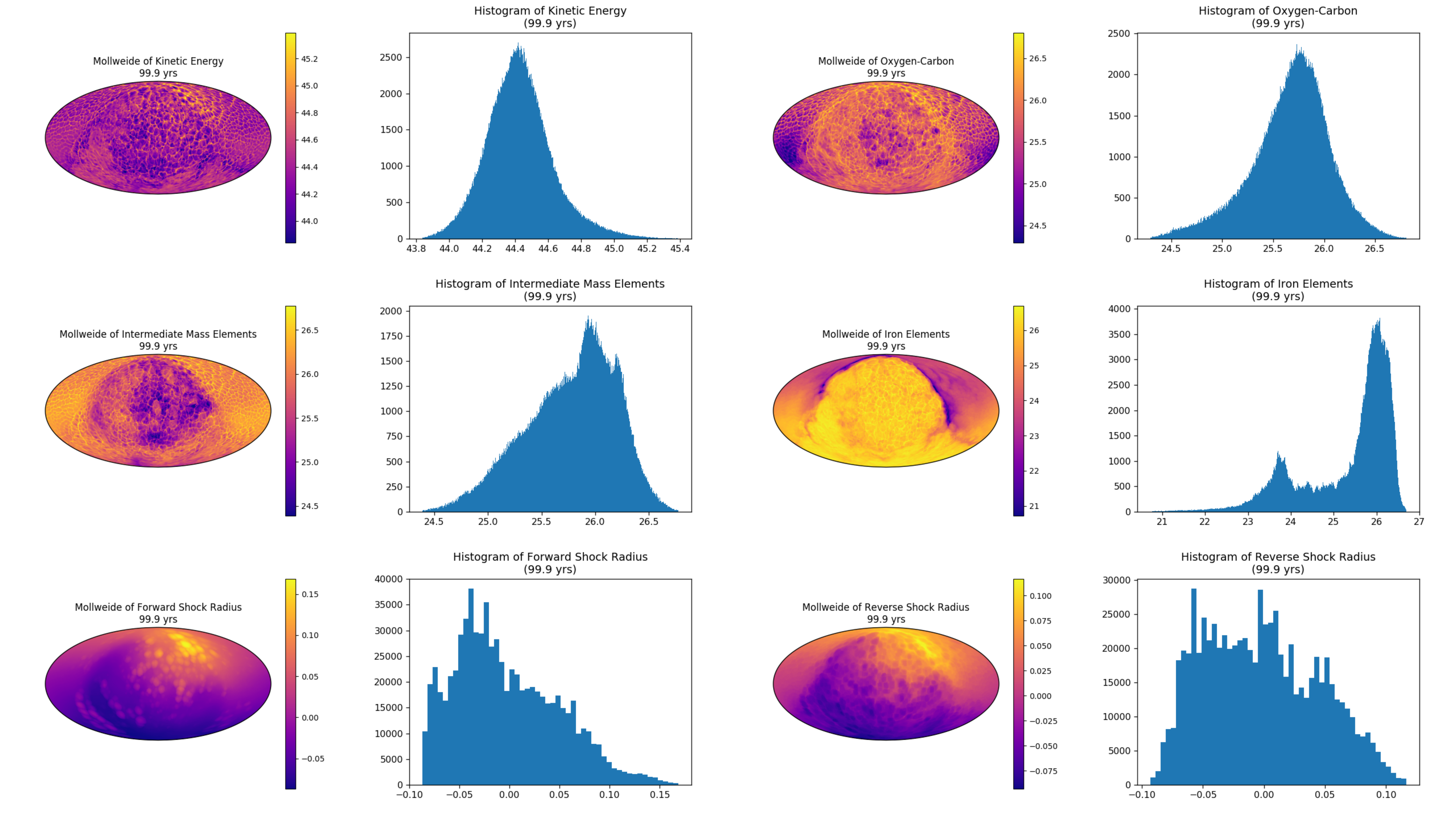}}
  \caption{As in Fig~\ref{13yr}, at an age of 100 years.
  This is about the age of G1.9+0.3.}
    \label{100yr}
\end{figure}

\begin{figure}
  \centerline{\includegraphics[width=17truecm]{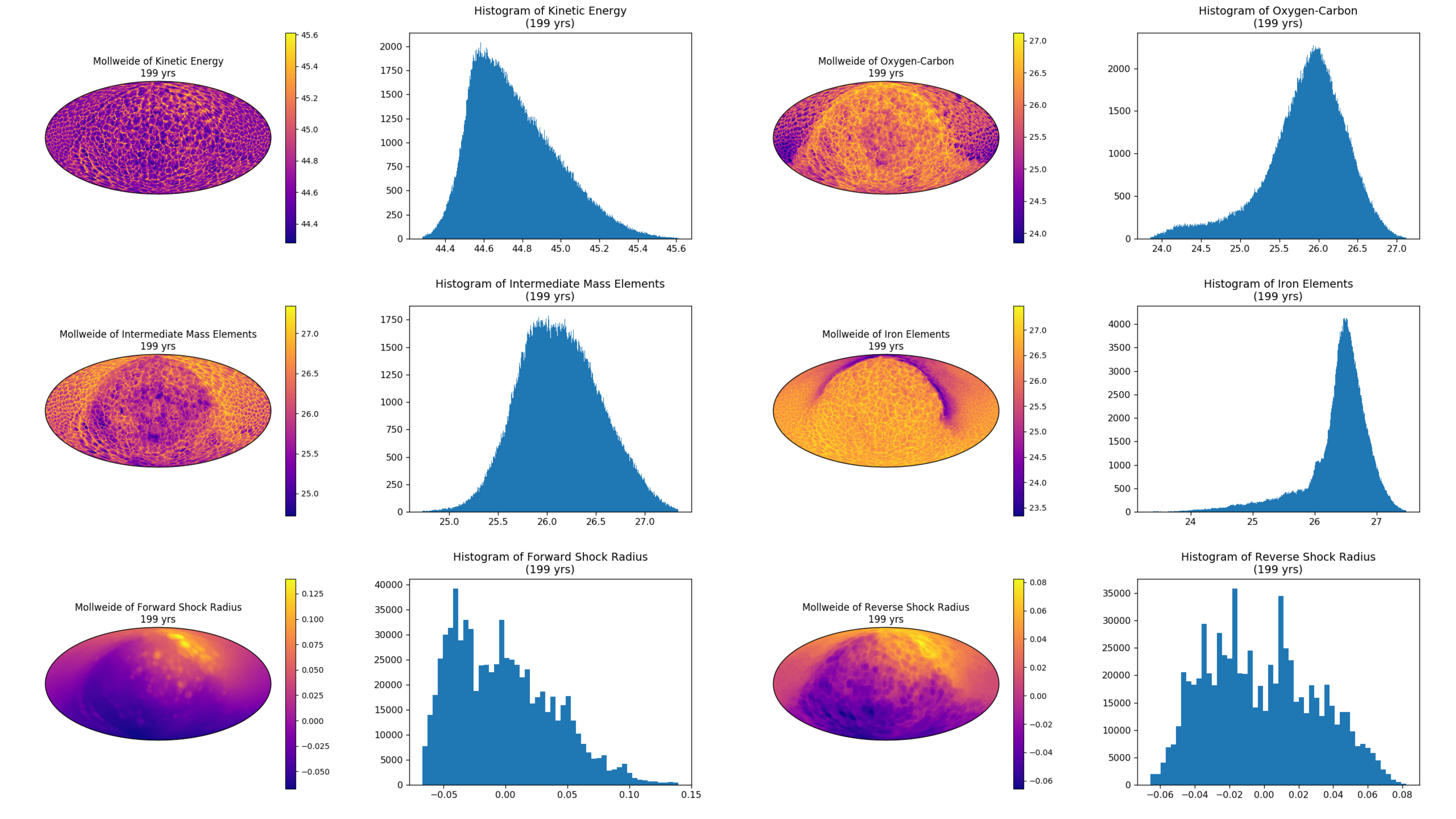}}
  \caption{As in Fig~\ref{13yr}, at an age of 200 years.}
    \label{200yr}
\end{figure}

\begin{figure}
  \centerline{\includegraphics[width=17truecm]{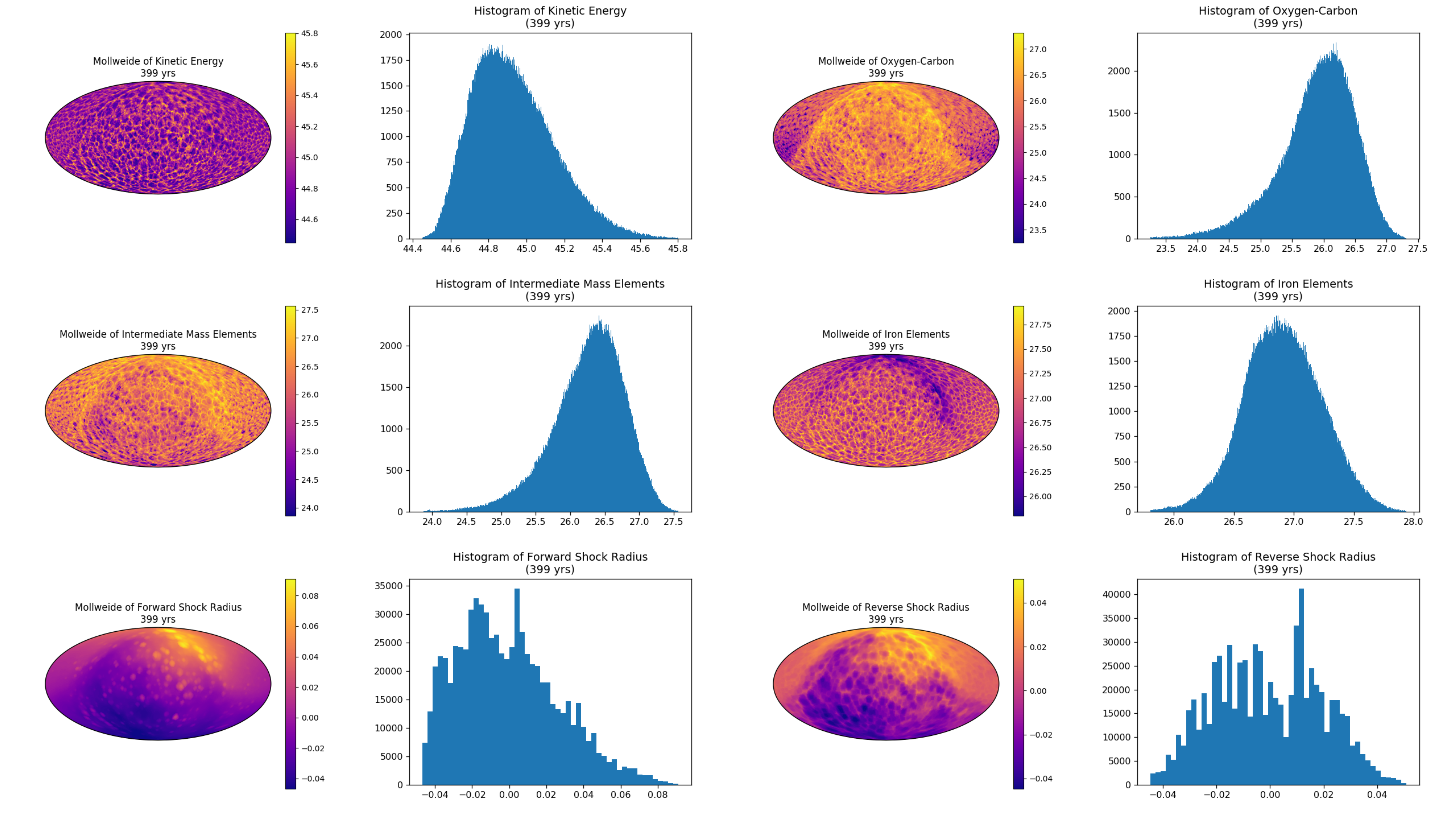}}
  \caption{As in Fig~\ref{13yr}, at an age of 400 years, roughly
  that of the remnants of Kepler's and Tycho's supernovae.}
    \label{400yr}
\end{figure}

\begin{figure}
  \centerline{\includegraphics[width=17truecm]{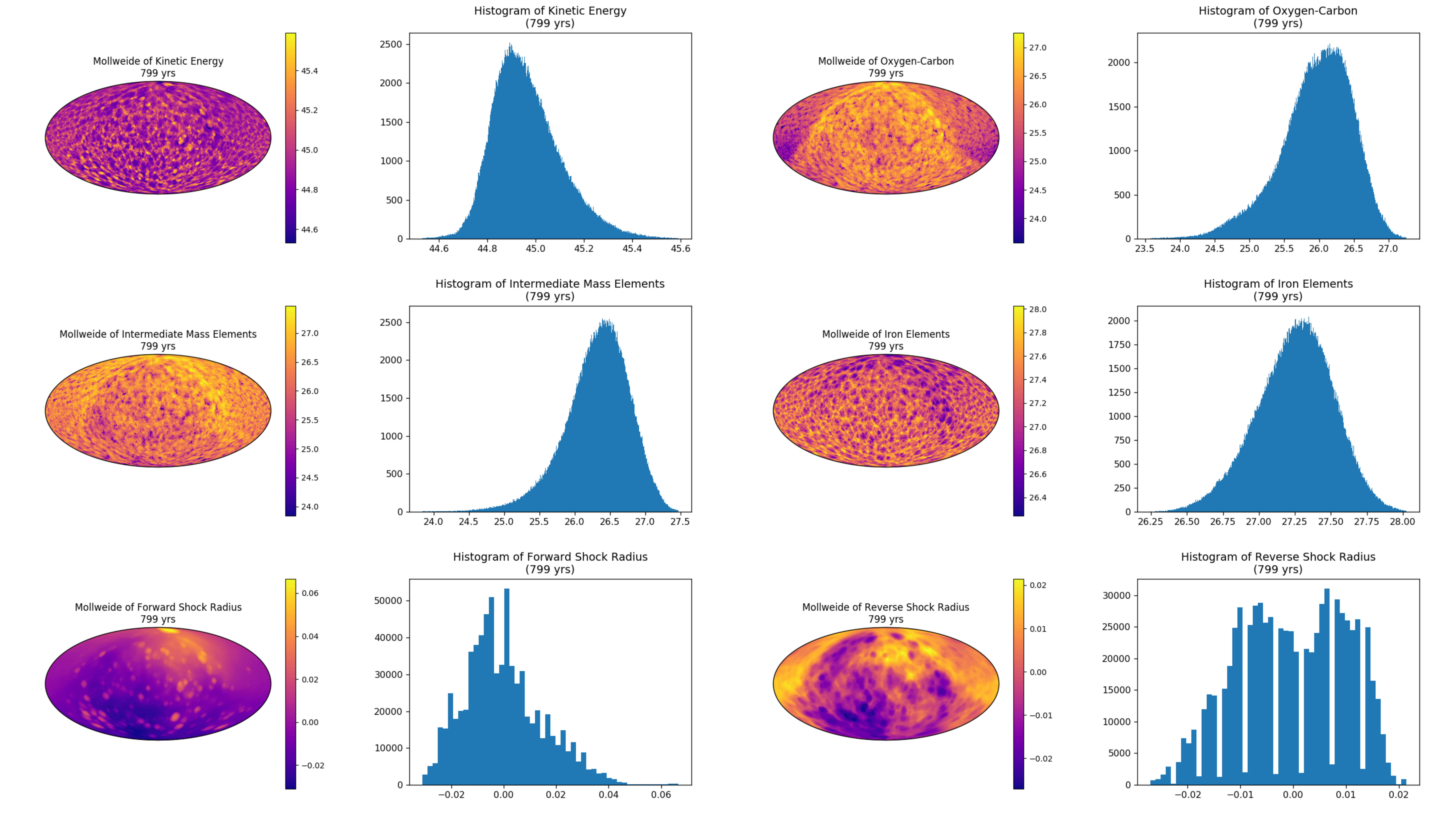}}
  \caption{As in Fig~\ref{13yr}, at an age of 800 years.}
    \label{800yr}
\end{figure}

\begin{figure}
  \centerline{\includegraphics[width=17truecm]{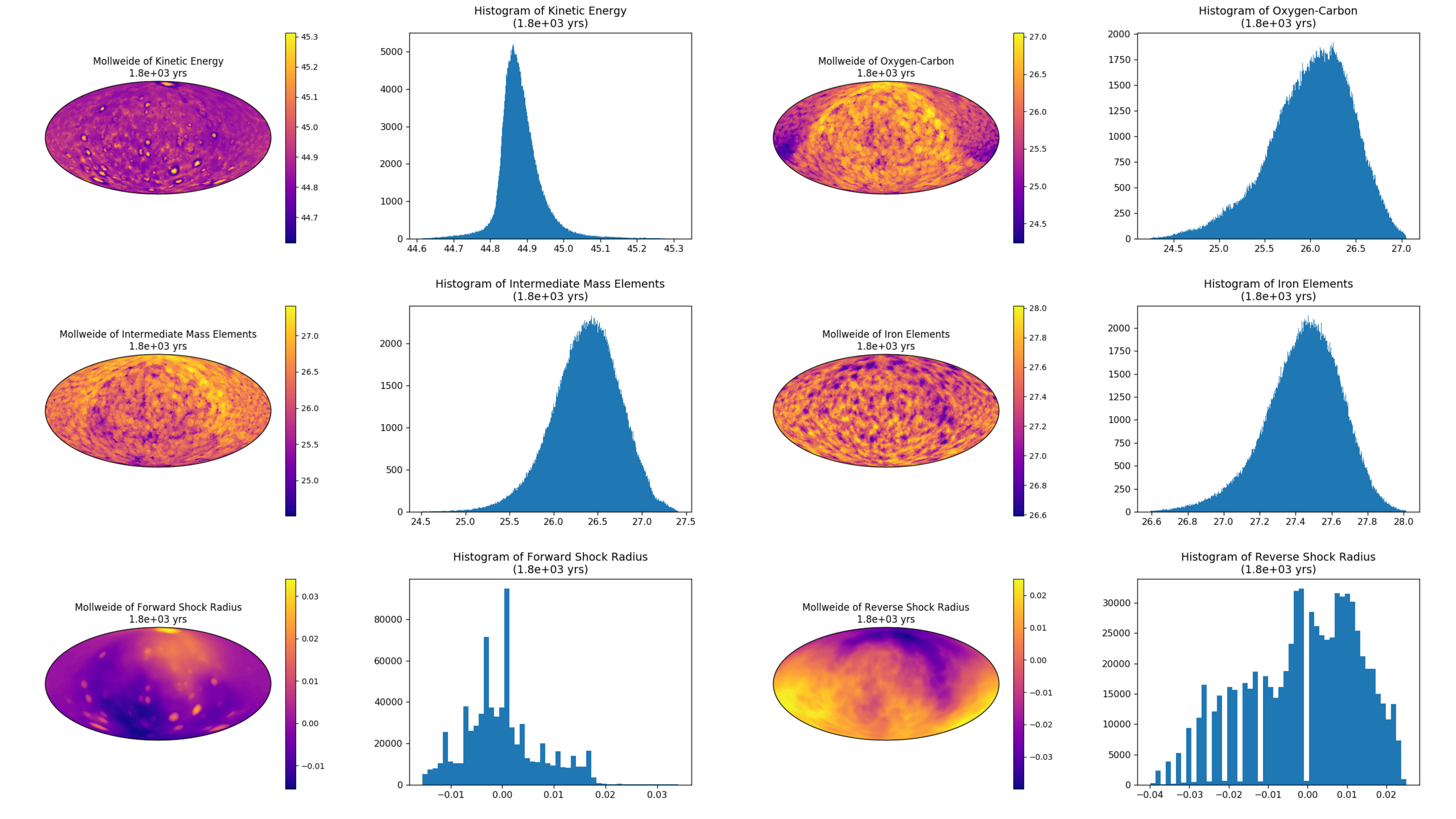}}
  \caption{As in Fig~\ref{13yr}, at an age of 1800 years,
  roughly the age of the likely SNR Ia RCW 86.}
    \label{1800yr}
\end{figure}

The model was evolved to an age of about 2000 yr, assuming a uniform
ambient medium of density $n_0 = 0.022$ cm$^{-3}$ ($\rho_0 = 5.08
\times 10^{-25}$ g cm$^{-3}$) \citep{carlton11}. Frames were extracted from the simulation
for ages of 13, 33, 75, 100, 200, 400, 800, and 1800 yr; the fourth
is about the estimated age of G1.9+0.3.  Montages of Mollweide
projections for various quantities are shown in Figures~\ref{13yr}
through \ref{1800yr}.

\begin{figure}
  \centerline{\includegraphics[width=12truecm]{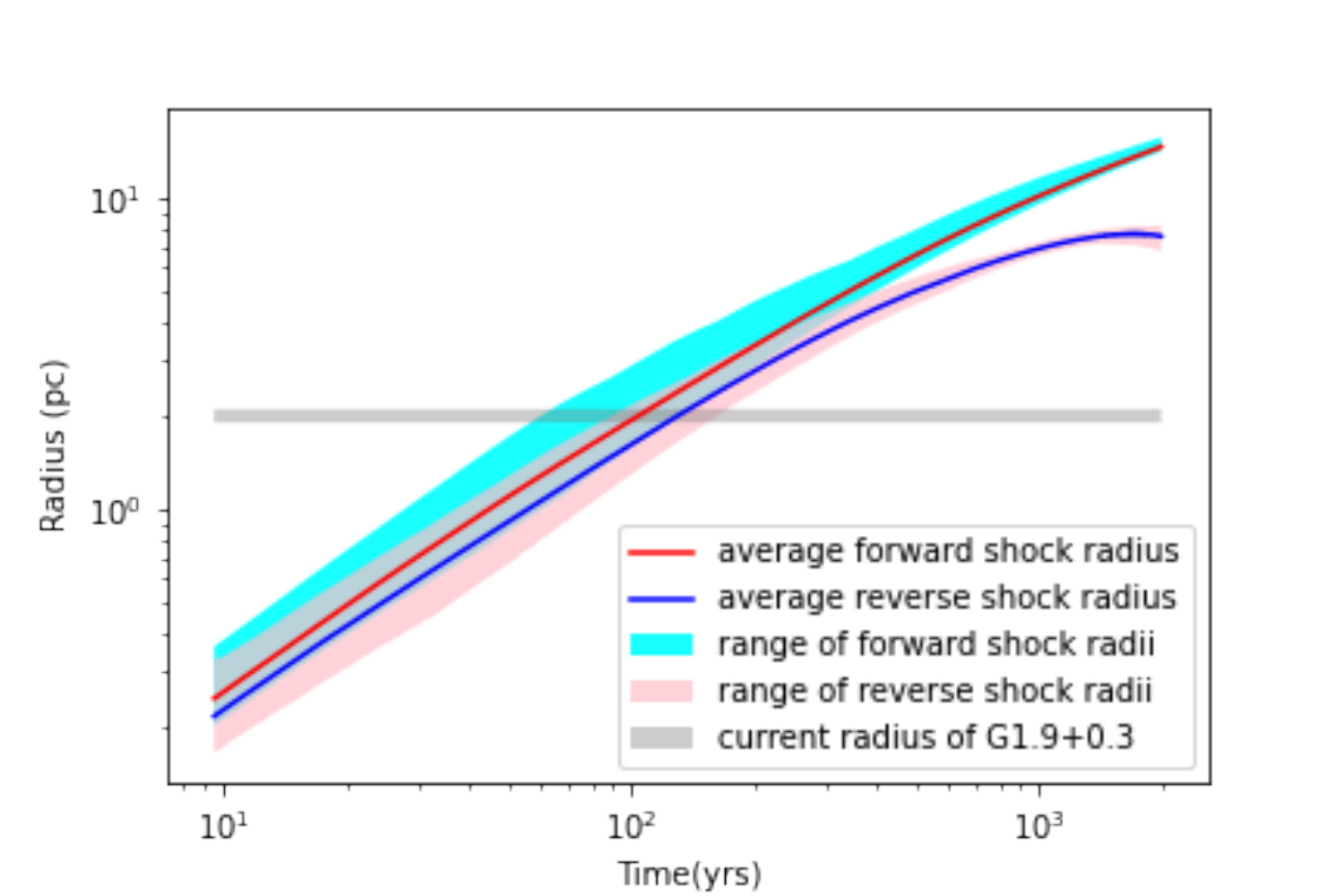}}
    \caption{Forward and reverse shock radii as a function of time, averaged
    over angles. The current observed shock radius, shown by the grey
    horizontal bar, is about 2 pc.}
  \label{radii}
  \end{figure}

\begin{figure}
  \centerline{\includegraphics[width=8truecm]{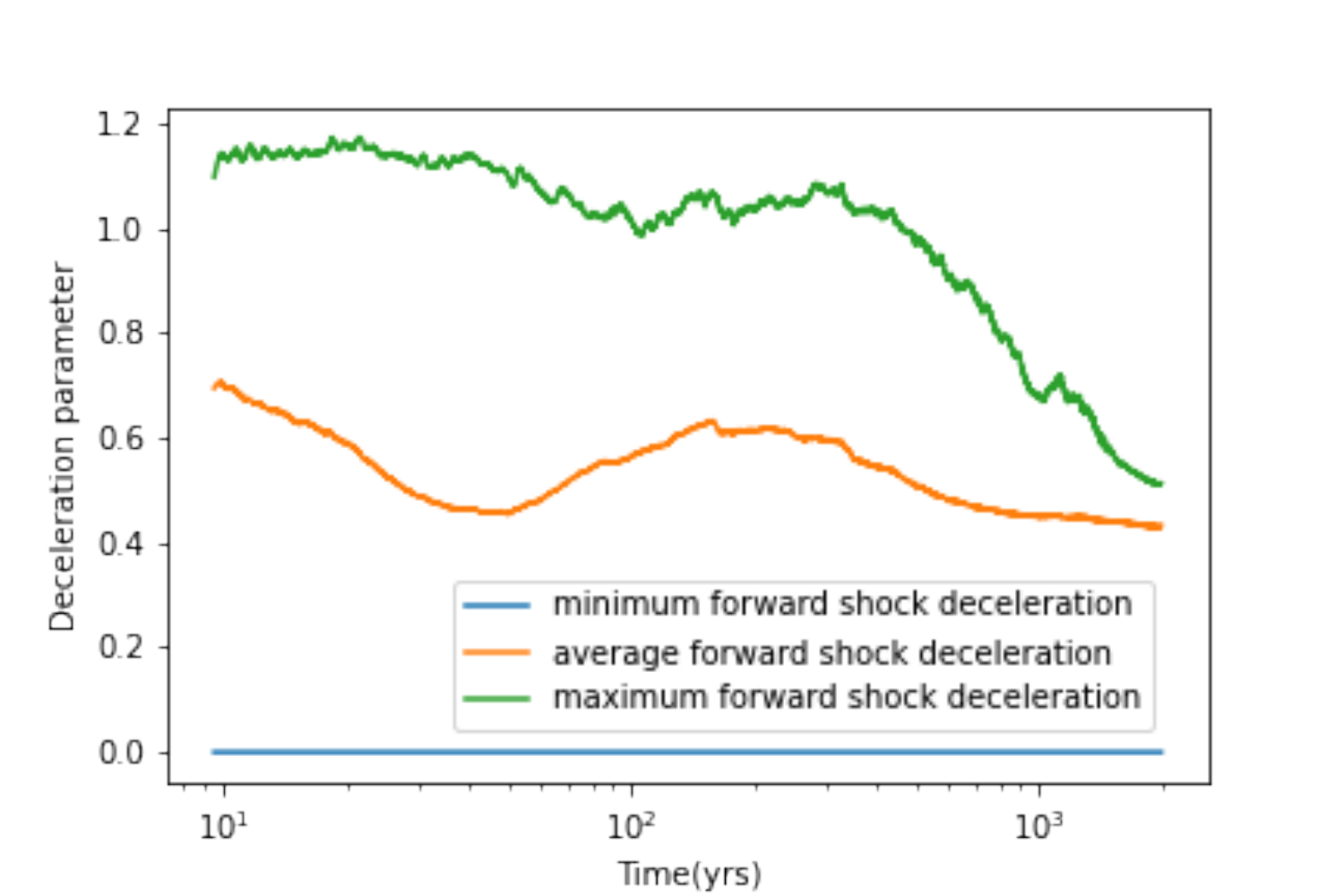}
    \hskip0.5truecm
    \includegraphics[width=8truecm]{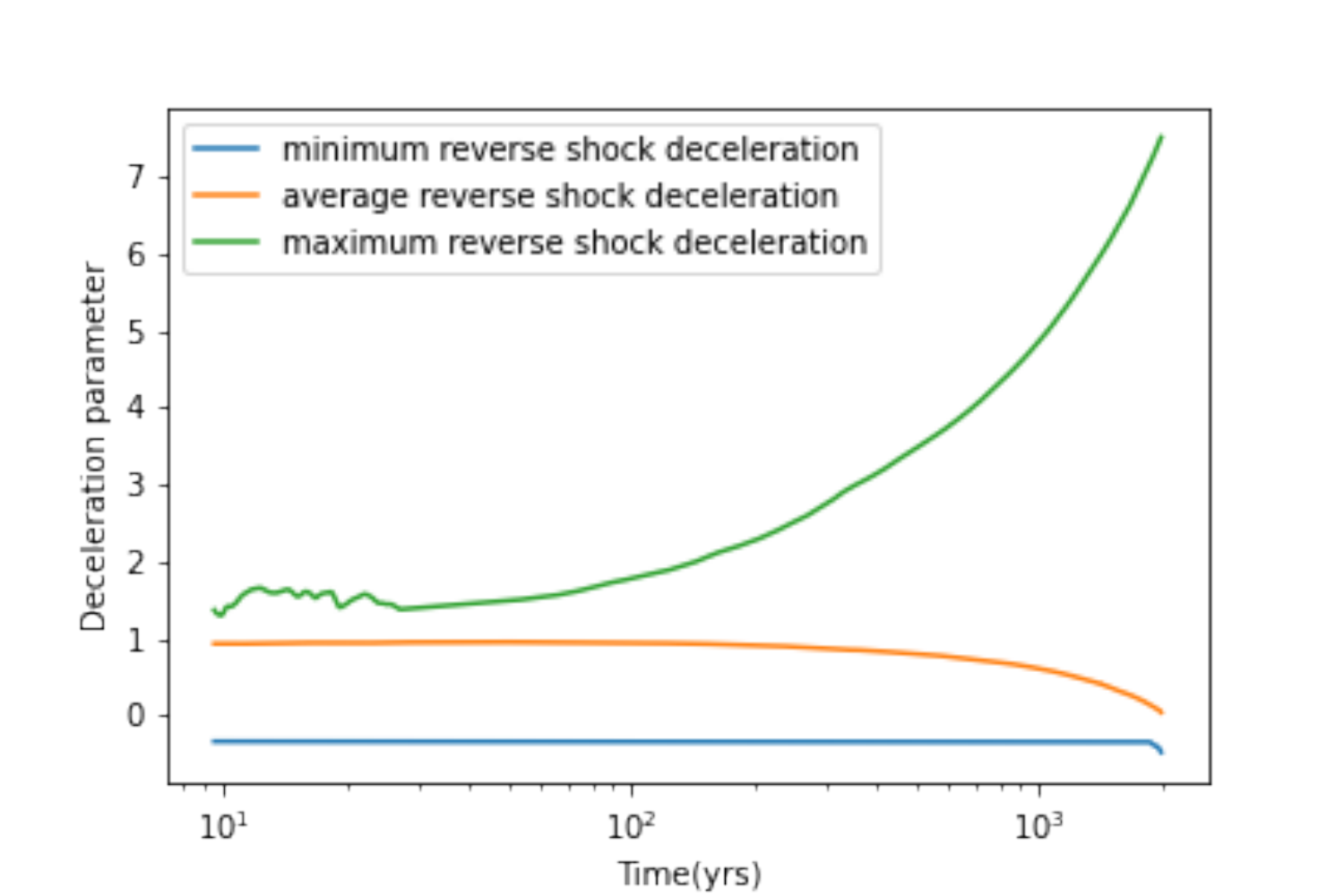}}
  \caption{Forward and reverse shock deceleration parameters as a
    function of time, averaged over angles.}
  \label{decel}
  \end{figure}

The evolution of the angle-averaged forward and reverse shock radii is
shown in Figure~\ref{radii}. The current observed shock radius is about
2 pc $\cong 6 \times 10^{18}$ cm, consistent with our age estimate of 100 yr. Figure~\ref{decel} shows the
forward-shock deceleration parameter $m \equiv d \log R/d \log t
\equiv vt/R$; while it shows variations, the deceleration of the
average shock radius evolves fairly smoothly toward the Sedov value of
0.4.  A 1D thin-shell analytic model fit to the remnant's current size and mean expansion rate \citep{carlton11} gave $m = 0.69$, between the average and maximum values we find here. At the end of the simulation, the total swept-up mass is about
12 $M_\odot$, by which time Sedov dynamics should be a good
approximation.  The reverse-shock deceleration drops to zero and
becomes negative as the reverse shock begins to move inward at an age
of around 2000 years.  It cannot be observed directly.

\begin{figure}
  \centerline{\includegraphics[width=8truecm]{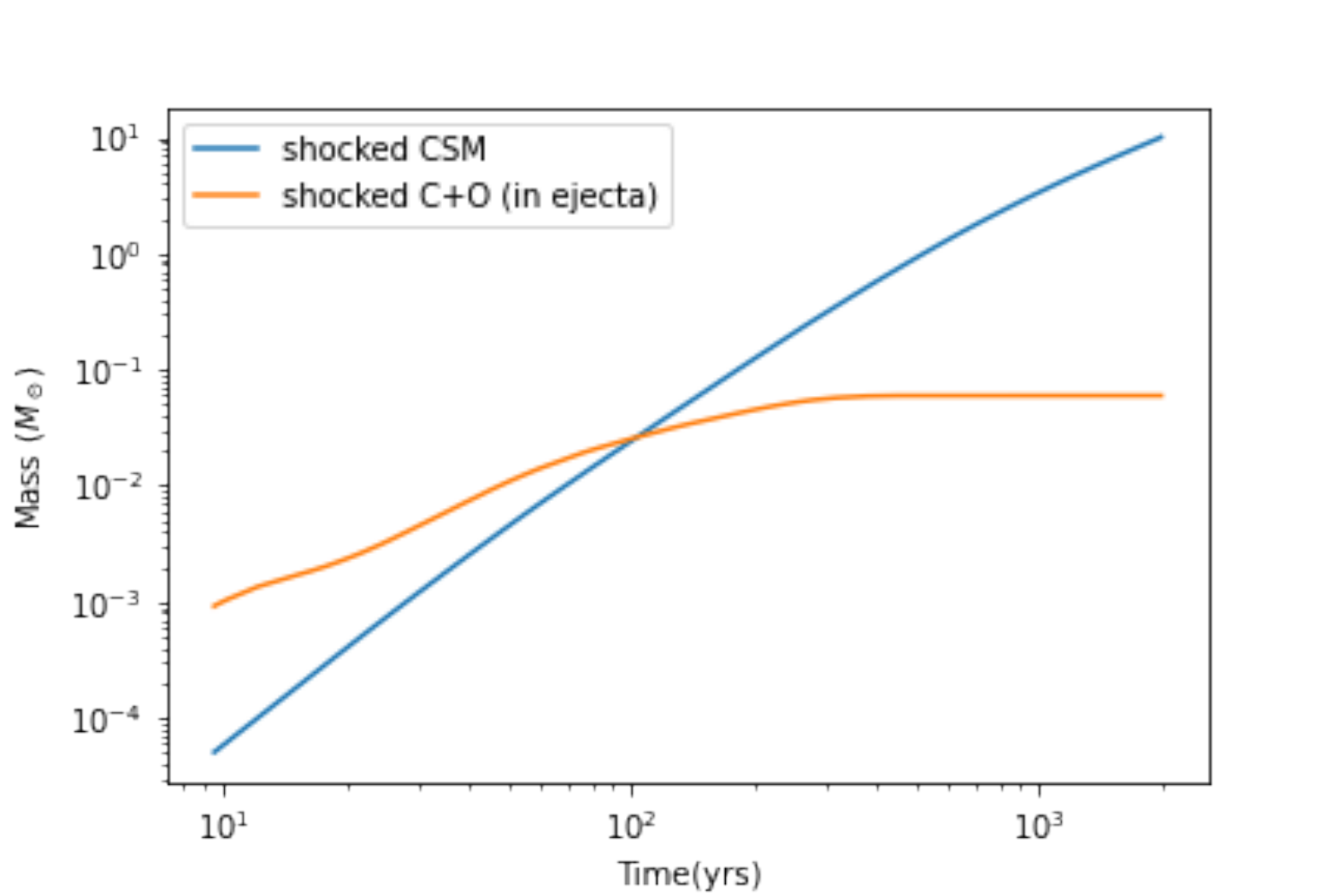}
    \hskip0.5truecm\includegraphics[width=8truecm]{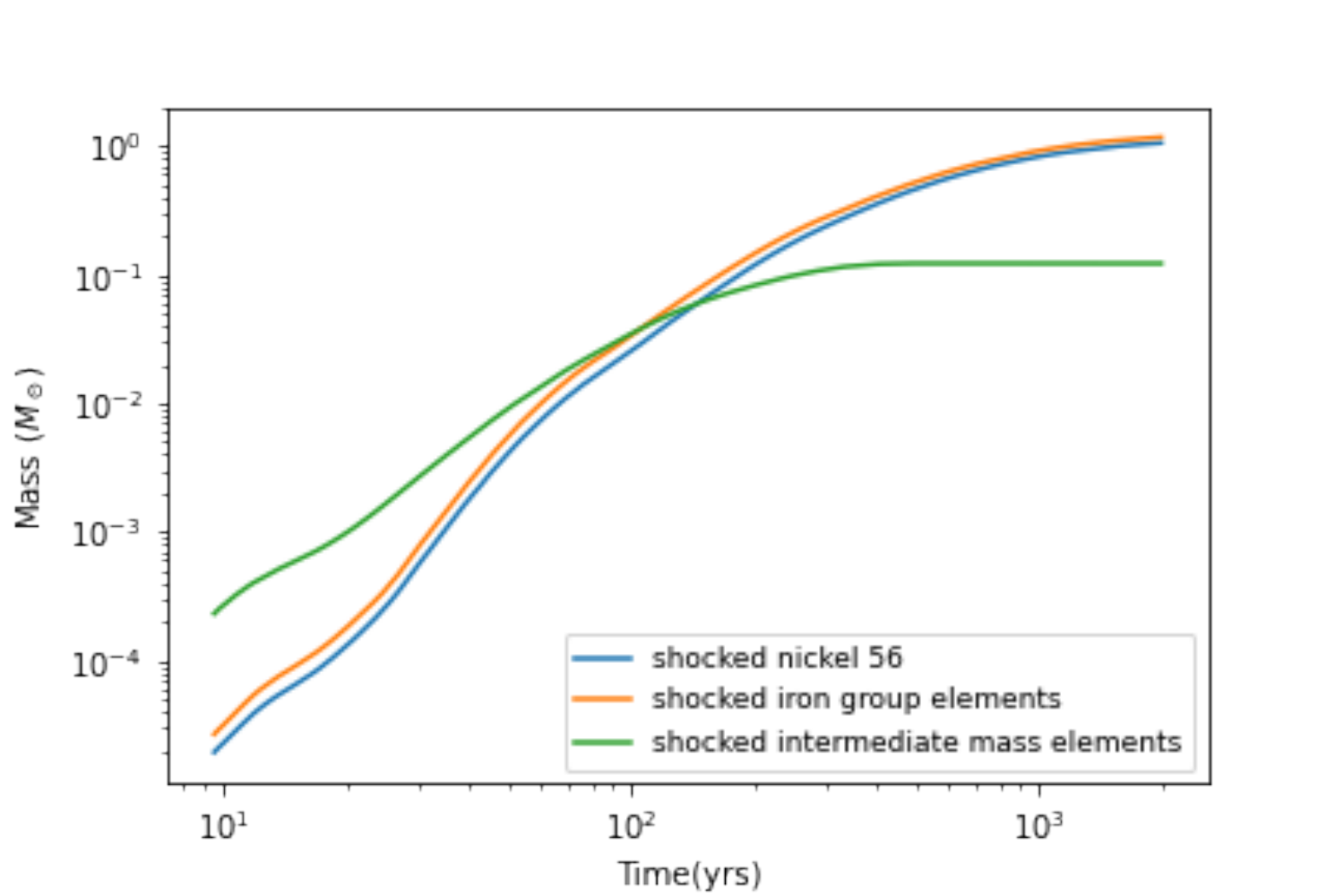}}
  \caption{Shocked masses of the various composition categories.}
  \label{Masses}
  \end{figure}

\begin{figure}
  \centerline{\includegraphics[width=8truecm]{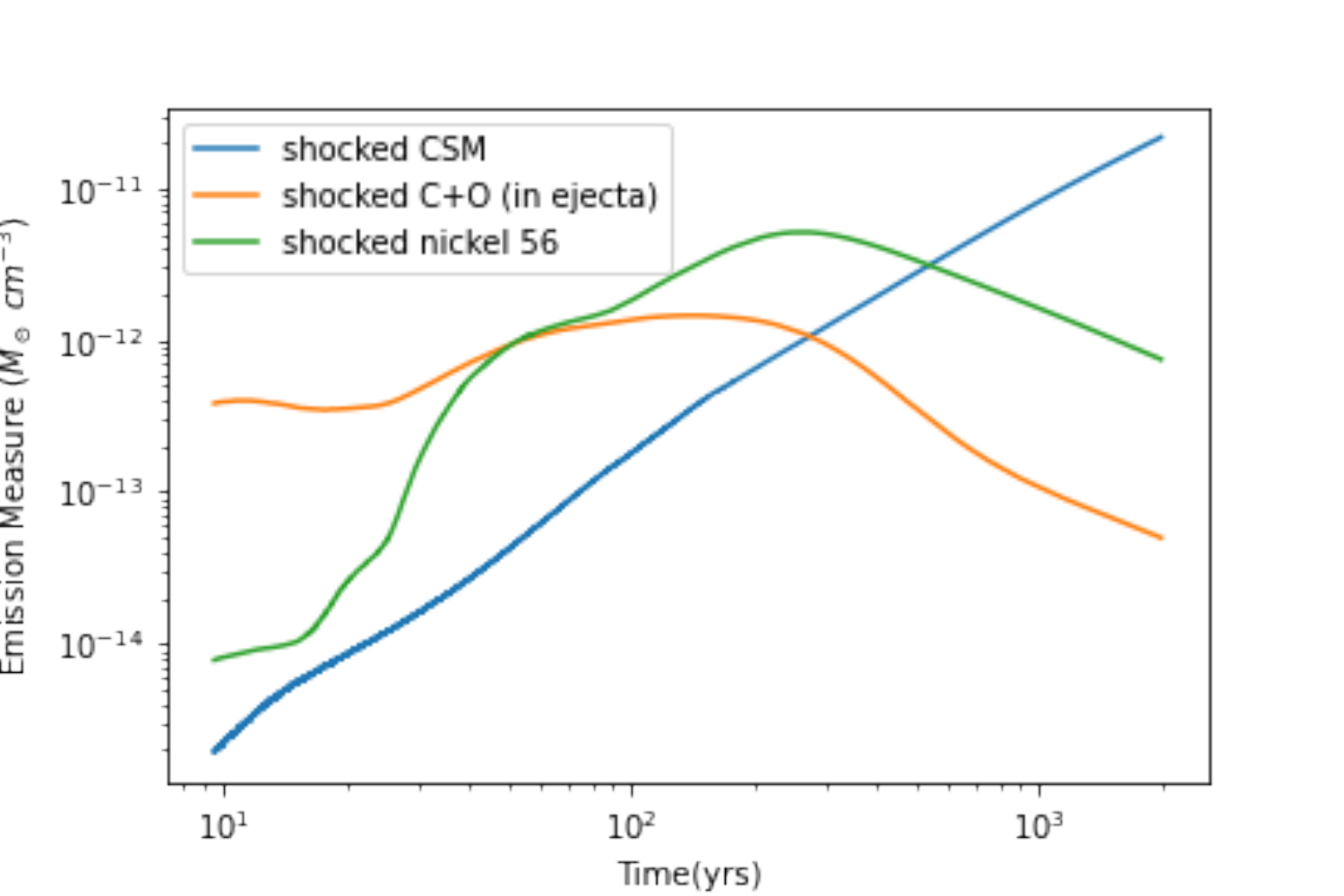}
    \hskip0.5truecm\includegraphics[width=8truecm]{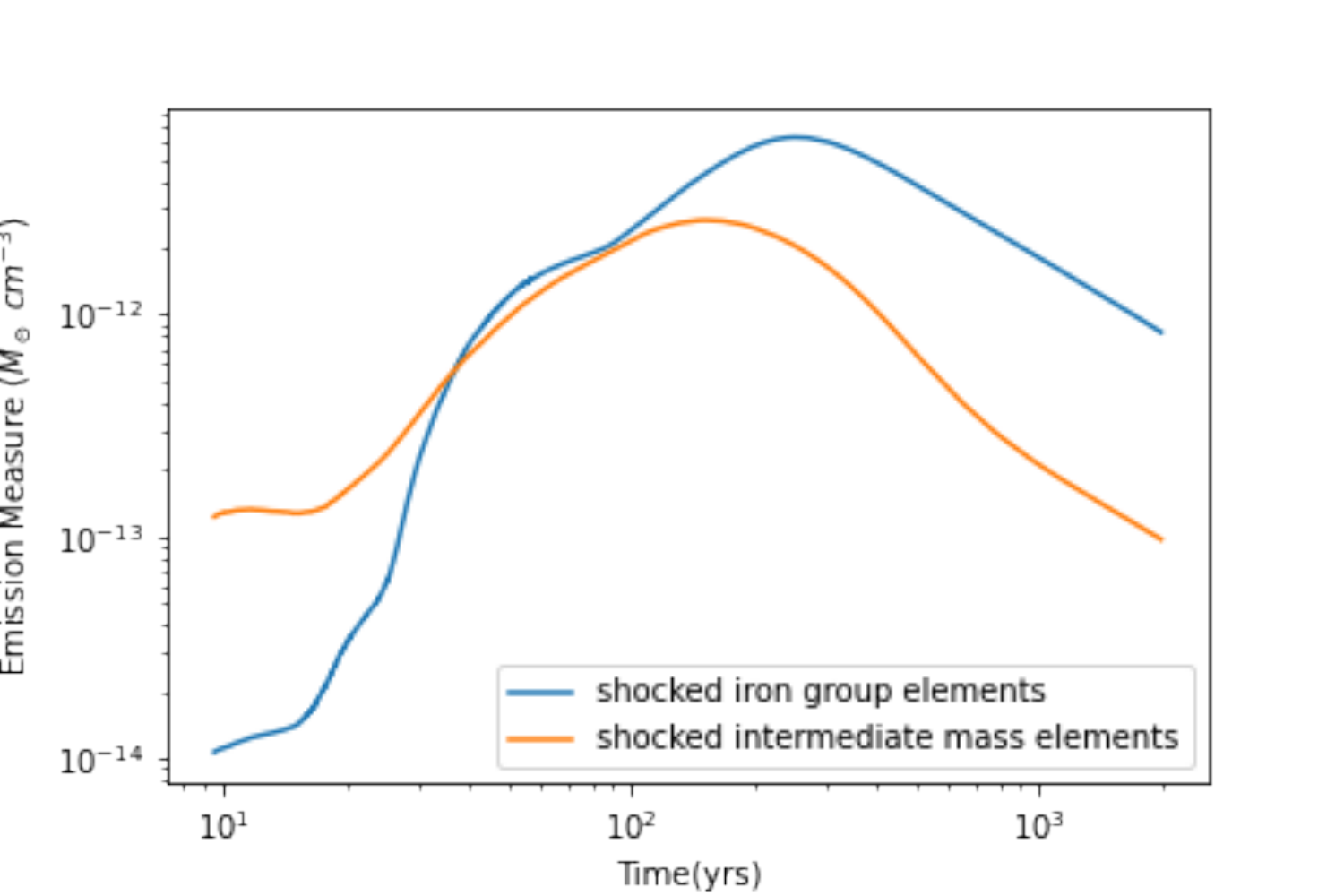}}
  \caption{Emission measures of the various composition categories (see
    text for definitions).}
  \label{EMs}
  \end{figure}

Shocked masses and emission measures (EMs) of the different composition
categories are shown in Figs.~\ref{Masses} and \ref{EMs}.  Here EM$ =
\int \rho f_i n_e dV$, where $\rho$ is the total density, $f_i$ are
the mass fractions of different species, and $n_e$ is the electron
density, calculated by approximating 11/13 electrons per amu of
CSM and 1/2 electron per amu of all other species.  We have assumed
full ionization of all species; observed EMs will reflect various confusing
effects, such as time-dependent ionization. IME's
in the simulation are fully shocked by about $10^{10}$ s, while
iron-group elements approach being fully shocked only near the end of the
simulation (Fig.~\ref{Masses}).  Of course, the shocked CSM mass continues to climb.
The shocked-CSM emission measure also continues to climb, but ejecta
EMs peak once all the masses are shocked, and then drop as $R_s^{-3}$.

The sequence of equal-area Mollweide plots (Figs.~\ref{13yr} through
\ref{1800yr}) shows the development of hydrodynamic instabilities,
most obvious near the end of the simulation.  At early stages, the
shocked masses are all quite small and show very pronounced spatial
variations, but as more material is shocked, these variations
decrease in amplitude.  Near the age of
\src\ (about 100 years; Figure~\ref{100yr}),
however, substantial asymmetry remains, on the relatively large scale
with which it was imprinted in the explosion.  Significant variations
in kinetic energy in different directions, still apparent out to 75
years, have mostly been smoothed away by 100 years, and are thereafter
dominated by hydrodynamic instabilities.  Comparing the sequence of
images shows where material is shocked.  Note particularly the
concentration of shocked iron in one hemisphere, surrounded by a
deficit.

\begin{figure}
    \centering
    {\includegraphics[width=14truecm]{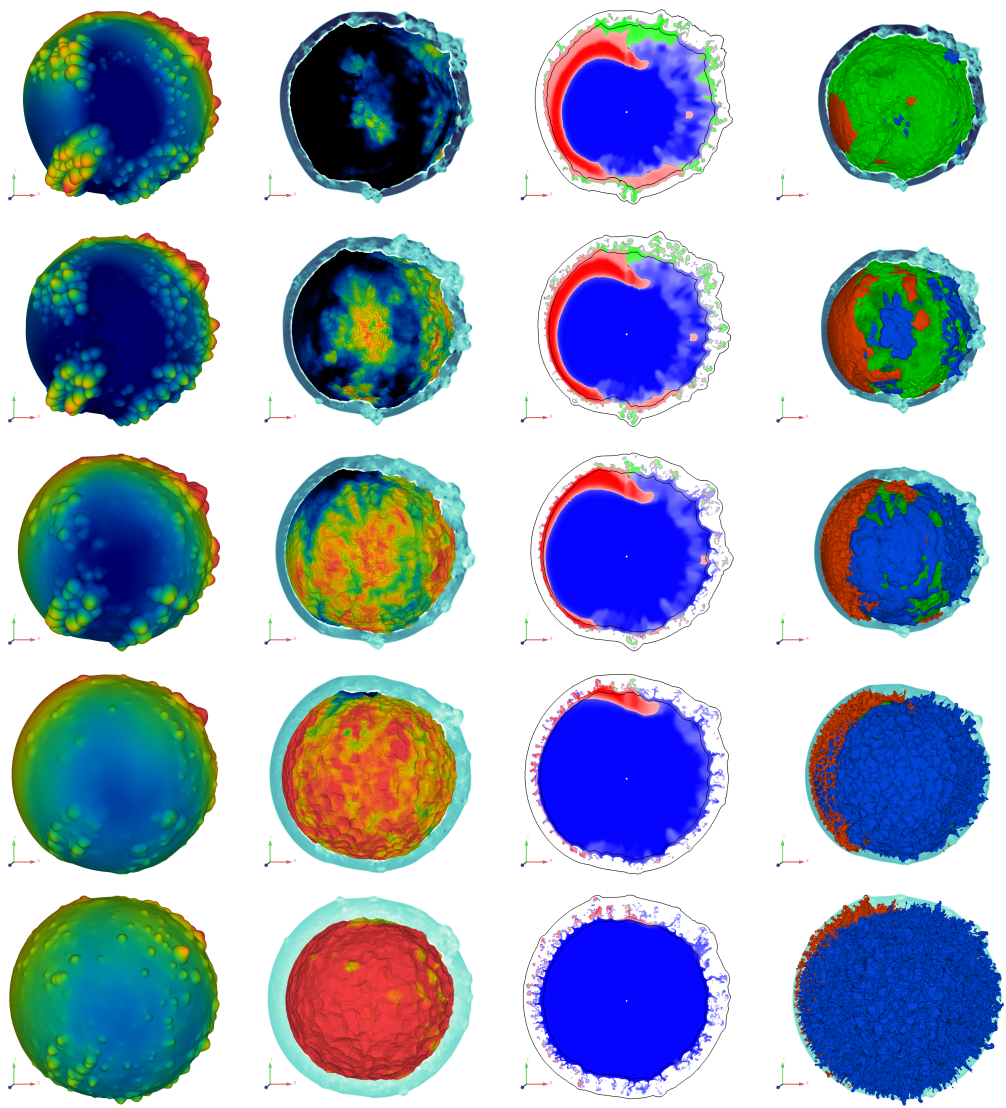}}
    \caption{\small{The structure of the SNR at ages of 25, 50, 100, 200, and 400 years (by row).
First column: surface of the forward shock colored by the deviation of the radial extent of the shock, with red protruding out beyond the average radius and blue lagging behind.  Second column:  three-dimensional surface of the reverse shock colored by mass fraction of iron from zero (black) to 100\% (red).  Light blue: gas pressure in a two-dimensional slice perpendicular to the view, illustrating the relationship between the reverse shock and the forward shock.  Third column: mass fraction in a two-dimensional slice through the center of the remnant.  IGEs in blue, IMEs in red, and C+O in green.  Black contours: forward and reverse shocks. Fourth column: mass fraction as three-dimensional surfaces.  IMEs (red) and IGEs (blue) shown as isosurfaces at 50\% mass fraction.  Green surface: reverse shock colored by mass fraction of C+O from zero (black) to 100\% (bright green). The pressure in the central plane shown as in Column 2.}}
    \label{imagearray}
\end{figure}

Fig.~\ref{imagearray} shows perspective views at ages of 25 to 400 yr, during which time the increasing smoothness of forward and
reverse shock surfaces can be observed.  Fe-group ejecta (blue in the last column of Fig.~\ref{imagearray}) are dominated by unburned C+O and IME's for the viewing angle shown, at the earlier times, but become more prominent with time.  

\section{Discussion}

\subsection{2D results}

The 2D simulations of the evolution of the model from \cite{kasen09}
show that for an ambient density that can reproduce the observed size
and expansion rate of G1.9+0.3, almost no iron-group elements will have 
been shocked after 100 years (Fig.~\ref{2Dmassfracs}), contrary to 
observations.  The asymmetries in the initial explosion model are 
substantial, but not sufficient to give rise to the overturn
(iron beyond silicon and sulfur) observed in G1.9+0.3.  Far greater
initial asymmetries will be required to reproduce these observed features.
The high-density model illustrates what is required to obtain sufficient
shocked iron; it is completely inconsistent with the observed size and
expansion rate of G1.9+0.3.  

\subsection{3D results}

Our 3D simulation begins with the most anisotropic of the suite of
models of \cite{seitenzahl13}.  With only three ignition points, all in
the same hemisphere, the
N3 model might be expected to show strong variations, and 
Fig.~\ref{initialmodel} substantiates this expectation.  While we focus 
on the evolution to
SNR stages, the appearance of such a supernova would clearly have
a substantial dependence on viewing angle.  Most iron is ejected
preferentially in a relatively small solid angle, roughly on the
opposite hemisphere from IMEs. The hemisphere opposite to the direction
of most iron injection is relatively smooth.  The viewing-angle
dependence of the calculated spectra and lightcurves for the N3 model
is discussed in some detail in \cite{sim13}.

As time advances (Figs.~\ref{13yr} to~\ref{75yr}), the amount of ejecta
passing through the reverse shock increases.  At an age of 13 years, 
relatively little iron is shocked.  The histogram shows that most
pixels have a small amount of iron, but a few reach much higher
values.  At 75 years, most pixels now have larger amounts of
shocked iron. (The overall mass of iron that is shocked is still
relatively small, about 0.01 $M_\odot$; see Fig.~\ref{Masses}.)

At the age of \src, Fig.~\ref{Masses} shows
comparable masses in all shocked components: CSM, ejecta C and O,
IMEs, and Fe-group elements.  In particular, only about $0.02\ M_\odot$
of $^{56}$Fe (originally $^{56}$Ni) has been shocked, though this is far larger
than the amount of shocked $^{56}$Fe found by an age of 100 yr in the
2D simulation.  The total shocked mass is about 0.1 $M_\odot$,
somewhat larger than in the 2D case.  This is also about three
times larger than
that estimated from observations of \src\ as described in Section~\ref{obs} above.
The preferential ejection of iron in some direction could be
identified with the stronger Fe features in the north rim of
\src\ \citep{borkowski13b}.  By the age of \src\ (Fig.~\ref{100yr}), a 
significant amount of shocked Fe can be seen to the
right of center in the images. 

As the remnant continues to evolve, the spread in all quantities decreases, 
as shown by the histograms.  At our earliest epoch of 13 years, different pixels vary in amount of iron by 8 orders of magnitude, while kinetic energy varies by a factor of about 30. At 100 years, the spread in KE is less than one order of magnitude. The nature of asymmetries varies as well.  Large-scale asymmetries visible in the initial model dominate to an age of 100 -- 200 years, at which point smaller-scale irregularities ascribable to Rayleigh-Taylor instabilities become apparent (see Fig.~\ref{200yr}).  By an age of 800 years, kinetic-energy values spread by only about a factor of 4, while shocked iron ranges over only about 20.  The reverse shock has become very smooth at late times (a small range of values), so the finite extent of
radial zones results in artifacts visible in the lower-right panels of Figures~\ref{800yr} and~\ref{1800yr}.  By the end of our simulation, roughly the age of the likely Type Ia SNR RCW 86 \citep{williams11}, unburned C and O still have a spatial distribution reminiscent of the initial model, while IMEs and IGEs are fairly smooth.  

An alternative view of the evolution of asymmetries is presented in Figure~\ref{imagearray},
showing isosurfaces of various quantities for different ages: forward and reverse shock
surfaces and mass fractions, and 2D slices through the remnant center of mass fractions,
for one particular viewing angle.  The same trends apparent in the Mollweide projections
can be seen here.

Our simulations show that the imprint of a highly asymmetric thermonuclear explosion is clearly detectable in the remnant for at least 500--1000 years, though it decreases in prominence as evolutionary asymmetries due to instabilities take over.  Of course, we have assumed a uniform external medium; when strongly asymmetric ISM or CSM is present, morphological changes can be expected, though compositional asymmetries should not be affected.  The north rim of G1.9+0.3 expands five times more slowly than other parts of the
remnant \citep{borkowski17}, a circumstance likely due to asymmetric surroundings.

A similar simulation, focusing on the morphology of the forward and reverse shocks and contact discontinuity, was recently presented by \cite{ferrand21}, who evolved two models from the same suite of \cite{seitenzahl13}, N5 and N100 (that is, with five and one hundred ignition points, respectively).  The results of their N5 simulation bear resemblances to our N3 results, but the N3 model is even more asymmetric.   The shock locations we find do show a strong initial imprint of the asymmetries, though in percentage terms these are less dramatic than the compositional asymmetries (not presented in Ferrand et al.~2021).  Only three ignition points produce clear global-scale asymmetries, already distinct from the pattern (in the contact discontinuity) seen at an age of one year in the N5 simulation of \cite{ferrand21}.  Both models evolve toward greater symmetry with smaller-amplitude variations, as might be expected.  The interpretation of supernova-remnant observations can profit greatly from the investigation of initial supernova models in this way.

\section{Conclusions}

The youngest supernova remnants can be expected to exhibit the dynamics
of their birth event most clearly.  The youngest Galactic SNR, \src\, shows
remarkable levels of asymmetry in the location of ejecta of different
composition, with shocked iron at large radii, sometimes beyond IMEs
such as silicon and sulfur.  Our hydrodynamic simulations of the evolution
of two different asymmetric SN Ia models show that such pronounced
asymmetries at an age of about 100 years are not easily produced.
Our 2D model produces far too little shocked iron, and at too small
radii, to reproduce the {\sl Chandra} observations of \src.  This is perhaps
not surprising in view of the fairly symmetric distribution of iron in the initial
model (Fig.~\ref{2Ddist0}), but it is not obvious that evolutionary effects
are unable to produce adequate shocked iron, as our calculation demonstrates. The most
extreme 3D model, N3 from \cite{seitenzahl13}, shows that somewhat larger
ejecta masses are shocked by this age, possibly consistent with the
observations.  It must be borne in mind that ejecta are detected at all
only in a few regions in \src\, whose X-ray spectrum is dominated by
nonthermal synchrotron radiation.

The strong asymmetries observed in \src\ also argue against currently popular 
models of sub-Chandrasekhar (0.8 -- 1.1 $M_\odot$) detonations for SNe Ia, 
since pure detonations tend to produce little in the way of asymmetry 
\citep[e.g.,][]{sim10,chamulak12}, unless the companion also explodes 
\citep[e.g.,][]{pakmor12}.  In particular the high-velocity Fe required is 
difficult to produce unless progenitors possess a high-mass He layer, as in 
some double-detonation models \citep[e.g.,][]{woosley94,sim12}.  However, these
models also produce amounts of $^{44}$Ti in conflict with observations 
\citep{borkowski10, weinberger20}.

We exhibit the evolution of the initial asymmetry of our model in a
series of equal-area projections of the entire surface, which
illustrate the character of the asymmetries of model N3.  In
particular, a large plume of IME-rich material can be seen, directed
approximately opposite to iron-group material, though the angular
variation in initial expansion velocity and kinetic energy is less
pronounced.  After a few hundred years, the imprint of the initial
explosion asymmetry has faded, being replaced by the development of
hydrodynamic instabilities on smaller scales, and lessening, though
not eliminating, compositional traces of the explosion asymmetry.

For the particular case of \src, explosion asymmetries should still dominate,
although confused by apparently strongly asymmetric ambient medium.  
However, changes on timescales of decades can be seen in Figs.~\ref{75yr} 
and~\ref{100yr}. The most recent {\sl Chandra} observations were done in 
2019-2020, 13 years after the initial discovery.  We may be able to observe 
the fading of the explosion imprint as time goes on, though the dominant 
nonthermal component of the X-ray spectrum, which is still increasing 
\citep{borkowski17}, continues to impede more detailed compositional 
studies.  In any case, \src\ challenges models of Type Ia supernovae and 
its continuing study should be highly informative for the development of 
such models.  In general, our results illustrate the value of detailed 
observations of the dynamics of the youngest SNRs and their power to 
constrain supernova explosion models.

\begin{acknowledgments}

This work was supported by the National Science Foundation through
grant AST-1062736 to NCSU's REU program, and by NASA through the {\sl
  Chandra} Guest Observer program (GO1-12098A and B).  A preliminary
version of this work was presented at the 2011 meeting of the
High-Energy Astrophysics Division of the American Astronomical Society
\citep{griffeth11}.
\end{acknowledgments}

\bibliography{snr}
\bibliographystyle{aasjournal}
\end{document}